\documentstyle[12pt,epsf]{article}
\newcommand{\beq}{\begin{equation}}
\newcommand{\eeq}{\end{equation}}
\newcommand{\bea}{\begin{eqnarray}}
\newcommand{\eea}{\end{eqnarray}}
\newcommand{\barr}{\begin{array}}
\newcommand{\earr}{\end{array}}
\newcommand{\bmat}{\left( \begin{array}}
\newcommand{\emat}{\end{array} \right)}

\newcommand{\bit}{\begin{itemize}}
\newcommand{\eit}{\end{itemize}}
\newcommand{\ul}{\underline}
\newcommand{\half}{\frac{1}{2}}
\newcommand{\third}{\frac{1}{3}}

\newcommand{\id}{{\sf{'}}{\hspace{-0.8mm}|\hspace{-0.6mm}|}\hspace{-1.6mm}
\rule[-1mm] {1.5mm}{.1mm}\hspace{-1.1mm}\rule[3.0mm]{.65mm}{.1mm}}
\newcommand{\B}{{^8}\!B}
\newcommand{\Be}{{^7}\!Be}

\newcommand{\Cl}{{^{37}}\!Cl}
\newcommand{\Ga}{{^{71}}\!Ga}
\newcommand{\res}{(N_e)^{res}}
\newcommand{\ncr}{(N_e)^{cr}}
\newcommand{\nmax}{(N_e)_{max}}
\newcommand{\Psurv}{\bra P(\nu_e \ra \nu_e) \ket}
\newcommand{\bra}{\langle}
\newcommand{\ket}{\rangle}

\newcommand{\nue}{\nu_e}
\newcommand{\nm}{\nu_{\mu}}
\newcommand{\nt}{\nu_{\tau}}
\newcommand{\temu}{{\theta}_{12}}
\newcommand{\tetau}{\theta_{13}}
\newcommand{\tmt}{\theta_{23}}
\newcommand{\th}{\theta}
\newcommand{\thm}{\theta^m}
\newcommand{\ra}{\rightarrow}
\newcommand{\impl}{\Longrightarrow}
\newcommand{\df}{\Delta f}
\newcommand{\dF}{\Delta F}
\newcommand{\dm}{\Delta m^2}

\begin{document}
\begin{titlepage}
\title{Three--Flavor Gravitationally--Induced Neutrino Oscillations
and the Solar Neutrino Problem}
\vspace{10pt}
\author{J.\ R.\ Mureika\thanks{newt@avatar.uwaterloo.ca} ~\\
and \\
\\
R.\ B.\ Mann\thanks{mann@avatar.uwaterloo.ca}\\
\\
{\it Department of Physics} \\
{\it University of Waterloo} \\
{\it Waterloo, Ontario N2L 3G1  Canada}
} 
\maketitle

\begin{abstract}
 
Some implications of the proposal that flavor nondiagonal
couplings of neutrinos to gravity might resolve the solar neutrino
problem are considered in the context of three neutrino flavors.
The two--flavor model is discussed as a limiting case of the full
three-generation mechanism, and the behavior of the $\nue$ survival
probability for various values of the three--flavor parameters is
studied.  Overlapping allowed SNU regions are obtained for the neutrinos 
which most likely contribute to the observed solar neutrino 
deficiency, and the effects of the addition of a third flavor are discussed.  
The extension to a three--generation framework
is found to yield a greater allowed region of parameter space, suggesting
that gravitationally--induced neutrino oscillations remain a viable
explanation of the Solar Neutrino Problem.

\vspace{20pt}
WATPHYS TH-96/02
\end{abstract}
\end{titlepage}

\section{Introduction}

	For three decades, resolution of the Solar Neutrino
Problem (SNP) has eluded experimental
and theoretical particle physicists alike.  There are now four
independent experiments for which the observed incident $\nue$ 
fluxes and events \cite{expts} have consistently been less than
half the predicted rate, according to the various existing
standard solar models \cite{bah3}.  

	A myriad of adjustments to existing astrophysical and
particle models have been made in an attempt to resolve the SNP.
These range from alterations of various
solar models to changes in the fundamental properties of the
neutrinos themselves, typically by modifying the
physics of neutrinos so as to permit neutrino oscillations.
The most popular of these latter proposals is
the MSW (oscillation) Mechanism \cite{msw1,wolf}, in which 
neutrinos (like quarks) are
assumed to possess distinct non--trivial flavor {\em and} mass 
eigenbases in which their state evolution may be described.
These are related by a unitary rotation matrix, and it is from this
relation that the flavor--oscillation behavior arises in the
equations of motion.

As there is at present
no direct experimental evidence that neutrinos have mass
it is worthwhile considering other possible mechanisms which could
give rise to neutrino oscillations. One such possibility was proposed
several years ago by Gasperini \cite{gasp1}, who noted that if
each flavor of neutrino $\nu_i$ possesses a different
gravitational coupling $G_i$, then neutrino oscillations would
be induced by gravitational effects. This mechanism
violates the  (Einstein) Equivalence Principle in the neutrino sector
and has recently been dubbed the VEP mechanism \cite{bah1}. However
it admits neutrino oscillations for neutrinos of degenerate or zero
mass, and so stands out as an interesting alternative to the mass
oscillation mechanism\footnote{Both mechanisms are not 
mutually exclusive.  There have been several papers which have
considered a massive--VEP oscillation mechanism, where the neutrinos
possess {\em three} distinct eigenbases (electroweak, massive,
and gravitational) \cite{gasp1,min1}.}.

To date, all analyses \cite{gasp1,bah1,pant1,mal1,ut1} of 
solar  neutrino data utilizing the VEP model have been performed with only
two flavors of neutrinos.  There is some debate as
to whether or not the allowed (two--dimensional) parameter space is large 
enough to validate a solution therein \cite{mal1,bah1}. More recently,
it has been pointed out \cite{ut1} that an improvement on the recent 
LSND experiments \cite{LSND}
have the potential for significantly constraining (or even ruling out)
the VEP mechanism in conjunction with solar and atmospheric
data.  

Motivated by the above, we consider in this paper 
a study of the realistic three--flavor VEP mechanism
in the context of the SNP. We analyze current solar 
neutrino data to obtain the allowed regions of 
parameter space in the 3-generation VEP mechanism.  
To simplify the analysis, we shall assume that the mixing matrix 
transforming the gravitational eigenstates to weak ones is real
({\it i.e.} no CP-violating parameter in this sector).
We show how the two--flavor ``limit'' may indeed
be recovered for various values of mixing angles, and study the
effects of hierarchy breaking in the violation parameters of the
$\nm$ and $\nt$.  We compute the $\nue$--survival probability surfaces for
the three flavor case. We conclude with a discussion of the viability
of the three flavor VEP mechanism as a solution to the SNP.

\section{Three--Flavor VEP Oscillation Formalism}

  For $N_{g}$ flavors of massless neutrinos, the Lagrangian may
be written
\beq
{\cal L} = i \sum_{i=1}^{N_{g}} e(G_i)\, \overline{\nu}_{G_i} e^{a \mu}(G_i)\,
\gamma_a D_\mu(G_i)\, \nu_{G_i} + {electroweak\; interactions}~.
\eeq
As usual, $e^a_\mu$ are the veirbeins, $\gamma_a$ the Dirac matrices,
and $e(G_i) \equiv det\{e^a_\mu(G_i)\}$.
Here, $G_i$ are the flavor-dependent gravitational coupling constants, 
and we denote 
\beq
G_i \equiv (1+f_i) G ~,
\eeq
where $f_i \ll 1$ are the VEP parameters which will play a
role similar to that of the neutrino masses in the more conventional
MSW mechanism.  Here $G$ is Newton's constant as defined by the coupling
of baryonic/photonic matter to gravity. 
	
We define the neutrino eigenstates as
\begin{equation}
|\nu_W \ket = \left( {\begin{array}{c} \nue \\ \nm \\ \nt
\end{array}} \right) ~,~|\nu_W \ket = V_3 \, |\nu_G \ket ~,
\label{eigens}
\end{equation}
where $|\nu_W \ket$ is the weak eigenbasis and $|\nu_G \ket$ the gravitational
eigenbasis.  The matrix $V_3$ is the $3 \times 3$ leptonic analogue of the
CKM mixing matrix, which can be parametrized as
\beq
V_3 = e^{i \tmt \lambda_7} e^{-i \tetau \lambda_5} e^{i \temu \lambda_2}~,
\label{ckm}
\eeq
where $\th_{ij}$ are the
vacuum mixing angles between flavors $i$ and $j$, respectively. 
We have chosen $V_3$ to be real ($\delta = 0$), such that no
CP violation occurs in the neutrino sector. This is done for
simplicity, as there is nothing to rule out the possibility
of such a violation.  The $\lambda_{i}$ are the SU(3) generators, 
whose representation is chosen such that $V_3$ takes on the
form of the CKM (quark) mixing matrix.

Explicitly,
\begin{equation}
\small
V_3 =
\left( \begin{array}{ccc} c_{12} c_{13} & s_{12} c_{13} & s_{13} \\
-s_{12} c_{23} - c_{12} s_{13} s_{23} & c_{12} c_{23} - s_{12} s_{13} s_{23}
& c_{13} s_{23} \\ s_{12} s_{23} - c_{12} s_{13} c_{23} & -c_{12} s_{23} -
s_{12} s_{13} c_{23} & c_{13} c_{23} \end{array} \right)~,
\label{vmix}
\end{equation}
where
\begin{equation}
c_{ij} = \cos \th_{ij} ~,~ s_{ij} = \sin \th_{ij}~.
\end{equation}

For a spherically--symmetric curved spacetime with interior mass distribution
${\cal M}(r)$, we can write the linearized metric components \cite{mtw}
\bea
g_{tt}& =& 1 + \frac{2G{\cal M}(r)}{r} + o(G^2) \nonumber \\
- g_{rr}& =& 1 - \frac{2G{\cal M}(r)}{r} + o(G^2)~,
\label{metric}
\eea
in a heliocentric coordinate system.  The neutrino path is chosen as purely
radial.

Obtaining the appropriate veirbeins from (\ref{metric}),
we use the massless Dirac equation to obtain a variant of the Weyl equation,
\beq
i \frac{d}{dr} | \nu_G \ket = H | \nu_G \ket ~,
\label{weil}
\eeq
with Hamiltonian
\beq
H \equiv  -2E |\phi(r)| diag\{1+f_i\}~  + H_{\mbox{weak}},
\eeq
where $\phi(r)$ is the Newtonian potential
\beq
\phi(r) = - G \int\;d^3r' \frac{\rho(r')}{|r-r'|},
\eeq
with $\rho(r)$ the solar density.

Under the change of basis in (\ref{eigens}), the evolution equations become
\beq
\impl ~ i\frac{d}{dr} | \nu_W \ket = H' | \nu_W \ket~,\nonumber \\
H'  =  - 2E |\phi(r)| V_3 {\cal F} V_3^{-1} + {\cal A}(r)~,
\eeq
where $H'$ is 
off--diagonal in the violation parameters and mixing angles,
and
\begin{eqnarray}
{\cal F} & = & \left( \begin{array}{ccc} 0 & 0 & 0 \\ 0 & \df_{21} & 0 \\
0 & 0 & \df_{31} \end{array} \right) \nonumber \\
{\cal A}(r) & = & \left( \begin{array}{ccc} \sqrt{2} G_F N_e(r)  & 0 & 0 \\
0 & 0 & 0 \\ 0 & 0 & 0 \end{array} \right) ~.
\end{eqnarray}
Note that ${\cal F}$ (or alternatively, $H'$) 
is defined modulo a total factor of unity $\id \cdot f_1$, which
merely results in an unobservable overall phase shift in the wavefunction.
Here, ${\cal A}(r)$ is the energy shift in the $\nue$
sector, due to charged--current interactions in $H_{\mbox{weak}}$, 
$N_e(r)$ is the density of electrons and $\phi(r)$ is the
radially--dependent solar potential.

It is convenient
to find a new {\em matter--enhanced} basis $|\tilde{\nu_M}\ket$ in 
which the Hamiltonian will once again be diagonal.  We define
\beq
i\frac{d}{dr} | \tilde{\nu_M} \ket = H'' |\tilde{\nu_M} \ket~,\nonumber \\
H''  = V_3^m H' (V_3^m)^{-1} ~,
\label{evol}
\eeq
The matter--enhanced mixing matrix 
$V_3^m$ in (\ref{evol}) required to re--diagonalize the evolution
equations is identical in form to (\ref{vmix}), after the substitution:
\begin{equation}
\th_{ij} \ra \thm_{ij}~,
\end{equation}
with $\thm_{ij}$ the matter--enhanced mixing angles. These are
somewhat more complicated in the three flavor case. For our purposes
we shall only require the angles $\th_{12}$ and $\th_{13}$:

\begin{eqnarray}
s_{m12}^2 & = & \frac{-(F_2^2 - \alpha F_2 + \beta) \dF_{31}}
{\dF_{32} (F_1^2 - \alpha F_1 + \beta) - \dF_{31} (F_2^2 - \alpha F_2
+ \beta)} \nonumber \\
s_{m13}^2 & = & \frac{F_3^2 - \alpha F_3 + \beta}{\dF_{31} \dF_{32}}~,
\nonumber \\
\label{matmix}
\end{eqnarray}
with
\begin{eqnarray}
F_1 & = & f_1 + \third \left\{ A - \sqrt{A^2 - 3B} S - \sqrt{3 (A^2-3B)
(1-S^2)} \right\} \nonumber \\
F_2 & = & f_1 + \third \left\{ A - \sqrt{A^2 - 3B}S + \sqrt{3 (A^2-3B)
(1-S^2)} \right\} \nonumber \\
F_3 & = & f_1 + \third \left\{ A + 2 \sqrt{A^2 - 3B}S \right\} ~,
\label{matmass}
\end{eqnarray}
and

\begin{eqnarray}
\alpha & = & f_3 c_{13}^2 + f_2 (c_{12}^2 c_{13}^2 + s_{13}^2) + f_1 (s_{12}^2
c_{13}^2 + s_{13}^2) \nonumber \\
\beta & = & f_3 c_{13}^2 (f_2 c_{12}^2 + f_1 s_{12}^2 ) + f_2 f_1
s_{13}^2 \nonumber \\
A & = & \df_{21} + \df_{31} + D \nonumber \\
B & = & \df_{21} \df_{31} + D \left[ \df_{31} (c_{12}^2 c_{13}^2
+s_{13}^2) \right] \nonumber \\
C & = & D \df_{21} \df_{31} c_{12}^2 c_{13}^2 \nonumber \\
S & = & \cos \left\{ \third \cos^{-1} \left( \frac{2A^3-9AB+27C}
{2 (A^2 - 3B)^{3/2}} \right) \right\}  \nonumber \\
D & = & \frac{\sqrt{2}}{2E|\phi(r)|} G_F N_e(r) ~.
\label{abc}
\end{eqnarray}
The MSW analogues of these parameters can be found in \cite{barg,zag},
and are related to the VEP terms in eqs.~(\ref{abc}) via the 
substitution
\beq
\frac{\dm_{ij}}{2E} \ra 2E|\phi(r)|\df_{ij}~.
\eeq
The trademark difference between the two oscillation mechanisms is
in their energy dependence: VEP effects are proportional to the neutrino
energy $E$, while in the MSW case they are inversely proportional
to $E$.  That is, we have

\bea
\Psurv_{VEP}& \sim& {\cal F}(E|\phi|\df_{ij}) \nonumber \\
\Psurv_{MSW}& \sim& {\cal F}(\frac{\dm_{ij}}{E}) 
\label{ecomp}
\eea
 This is clear from the
definition of $D$ in (\ref{abc}).  The authors of \cite{mal1}
suggest that an energy--resolution of less that $20\%$ in the
SNO and Superkamiokande detectors would be sufficient to 
desciriminate between VEP and MSW suppression.

Solving (\ref{evol}) for $|\nue\ket$,
the full (averaged) three--flavor $\nue$ survival probability can be 
written as
\begin{eqnarray}
\Psurv& =& \sum_{i,j=1}^{3} |(V_3)_{1i}|^2\; |(P_{LZ})_{ij}|^2\; 
|(V_3^m)_{1j}|^ 2 \nonumber \\
 & = & c_{m12}^2 c_{m13}^2 \left\{ (1-P_1)c_{12}^2
 c_{13}^2 + P_1 s_{12}^2 c_{13}^2 \right\} \nonumber \\
& & + s_{m12}^2 c_{m13}^2 \left\{ P_1 (1-P_2) c_{12}^2 c_{13}^2 + (1-P_1)(1-P_2)
s_{12}^2 c_{13}^2 + P_2 s_{13}^2 \right\} \nonumber \\
& & + s_{m13}^2 \left\{ P_1 P_2 c_{12}^2 c_{13}^2 + P_2 (1-P_1) s_{12}^2
s_{13}^2 +(1-P_2) s_{13}^2 \right\}~. \nonumber \\
{  }
\label{pthree}
\eea
      We may parameterize the non--zero probability
of level crossing by a matrix $P_{LZ}$ of the form \cite{bald}

\begin{eqnarray}
\small
\begin{array}{lll}
P_{LZ} &
= & \left( \begin{array}{ccc} 1 & 0 & 0 \\ 0 & \sqrt{1-P_2} &
\sqrt{P_2} \\ 0 & - \sqrt{P_2} & \sqrt{1-P_2} \end{array} \right)
\left( \begin{array}{ccc} \sqrt{1-P_1} & \sqrt{P_1} & 0 \\ \nonumber
- \sqrt{P_1} & \sqrt{1-P_1} & 0 \\ 0 & 0 & 1 \end{array} \right)
\end{array}
\\
\small
\begin{array}{lll}
 & = & \left( \begin{array}{ccc} \sqrt{1-P_1} & \sqrt{P_1}& 0 \\
- \sqrt{P_1 (1-P_2)} & \sqrt{(1-P_1)(1-P_2)} & \sqrt{P_2} \\
\sqrt{P_1 P_2} & - \sqrt{P_2 (1-P_2)} & \sqrt{1-P_2} \end{array} \right)
\end{array}
\end{eqnarray}
The terms $P_1$ and $P_2$ are the Landau--Zener jump probabilities for the
$\nue \ra \nm$ and $\nue \ra \nt$ transitions, respectively, and are
given by \cite{bah1}

\begin{equation}
P_{i} = \frac{e^{-\eta_i} - e^{-\xi_i}}{1 - e^{-\xi_i}} ~ ,
\end{equation}
where
\begin{equation}
\xi_i = 2 \pi \kappa_{i} \left( \frac{\cos 2 \theta_{1i}}{\sin^2 2 \theta_{1i}}
\right) ~,~ \eta_i = \frac{\pi}{2} \kappa_{i} (1- \tan^2 \theta_{1i}) ~.
\end{equation}
The functions $\kappa_i$ are the adiabaticity parameters for the $1i$-transition
s \cite{bah1},
\begin{equation}
\kappa_{i} =\frac{\sqrt{2}\: G_F\, (N_e)_{res}\: {\tan}^2{2 \theta_{1j}}}
{\left| \frac{1}{N_e}
\frac{dN_e}{dr} - \frac{1}{\phi}\frac{d\phi}{dr}\right|}  ~ ,
\label{adia}
\eeq
\noindent which are adiabatic for $\kappa_i \gg 1$, and highly
non--adiabatic for $\kappa_i \leq 1$.  

The definition of $\kappa$ given in eq.~\ref{adia} differs from its
MSW analogue in the addition of the potential differential term
$\frac{1}{\phi}\frac{d \phi}{dr}$.  However, this addition does not
account for a sizable variation in the value of the denominator.
The ambient solar electron density $N_e(r)$ varies by several orders
of magnitude in the interval $[0,R_{\odot}]$, which the potential
$\phi(r)$ increases by a less than a factor of 10.  Indeed, there
is some debate as to whether VEP oscillations are dominated by the
solar potential or by  the larger (and constant) potential of the
local supercluster \cite{pant2}.

We close this section by noting that the study of (\ref{pthree}) can be 
greatly simplified by considering certain special cases of $\thm_{1i}$, 
which are reflected in the interaction
between the $\nue$s and the surrounding matter.  Since the ambient solar
electron density $N_e(r)$ strictly decreases in the interval $r \in [0,
R_{\odot}]$, then neutrinos can either be created above or below their
resonance density.  If the former holds, then the $\nue$-creation density
$\ncr$ is interior to the resonance density $\res_{1i}$, and so the
$\nue$ will pass through a resonance as it propagates outward.  Conversely,
if $\ncr < \res_{1i}$ , the $\nue$ will never undergo resonance, and
will propagate as if in a vacuum\footnote{For first generation neutrinos, 
we shall set $f_1=0$, and subsequently  $f_j \ne 0, j \ne 1$.
We make this choice based on the results of ref. \cite{1gen}, 
which we shall take to indicate that electron neutrinos 
(antineutrinos) and photons couple to gravity with strength $G$.
Since the results we obtain are all dependent upon differences between
the EEP-violating parameters $f_i$, this is not too restrictive an 
assumption.}.  As a special case of this, we consider
only $\res > \nmax\,$, {\it i.e.} the $\nue$ resonance density exceeds
the maximal solar electron density.

The values of the matter--enhanced mixing angles $\thm_{ij}$ are
determined by the resonance conditions.  In the case $\ncr \leq \nmax
<< \res_{1i}$,
generally $\df_{i1} \gg D(r)$.  Conversely,
the $D$ term in (~\ref{abc}) will dominate in Eqs.~(\ref{matmix})
for  $\ncr >> \res_{1i}\,$.  
Thus,  it can be shown from Eqs.~(\ref{matmass})
and (\ref{abc}) that the matter mixing angles $\thm_{1i}$ 
will hence behave as follows in the two aforementioned scenarios:
\bea
\ncr \gg \res_{12,13}:& \thm_{12},\thm_{13} \ra \frac{\pi}{2} \label{cond1} \\
\res_{12,13} > \nmax:& \thm_{12},\thm_{13} \ra \temu, \tetau \label{cond2}~.
\eea
We caution, however, that (\ref{cond2}) holds {\em only if} there
is a natural hierarchy in the VEP parameters, i.e. $\df_{31} > \df_{21}$
As we will later demonstrate, the behavior is quite different for
a broken hierarchy (see 
Section~\ref{hier}). It should also be noted that the Landau--Zener
jumps $P_i$ vanish if the corresponding resonance density exceeds $\nmax$.

\section{The Two--Flavor Limit}
\label{twosnu}

Before proceeding to investigate the three-flavor model, we shall first
examine the two--flavor limit. 

We consider only the contributions of three 
types of solar neutrinos: $\B,~\Be$, and pp. These contributions turn
out to be the largest for all detectors currently in use.
For example, we have calculated the unreduced counting rate contribution
from the $\B$ and $\Be$ neutrinos to the $\Cl$ detector  to be $R_{\B} = 
6.24~$SNU and $R_{\Be} = 1.13~$SNU using data from \cite{bah2}. 
Combined, these
rates account for $93.3\%$ of the total SSM rate in Table~\ref{snussm}.
For the $\Ga$ detectors, the equivalent unreduced rates are $R_{pp} =
70.94~$SNU, $R_{\Be} = 34.40~$SNU, and $R_{\B} = 14.53~$SNU, comprising
$90.8\%$ of the total predicted rate.

   Two--generation oscillation analyses can be considered as
special limiting cases of the full three--flavor model.  Specifically,
they result when the $\nt$ (or $\nm$, for a broken hierarchy) totally
decouples from the rest of the
interactions, effectively reducing the $3\times 3$ matrix $V_3$
to a $2 \times 2$ matrix $R_2$ plus a sterile
neutrino.\footnote{We note that the third neutrino need not be the $\nt$,
as many theories incorporate a ``fourth generation'' neutrino, or
supermassive neutrino, and so forth.  Since we are considering a realistic
three--generation analysis based on the Standard Model, though,
we assume it to be the $\nt$.}

To do this, we must adhere to the following constraints,
$(i): \res_{13} > \nmax$, and (ii): $\tetau = 0$.  The former is 
satisfied for solar neutrinos of $E > 0.2\:$MeV if 
$E \df_{31} > 10^{-12}\:$ MeV, and so $\thm_{13} \ra \tetau$ and 
$P_2 \ra 0$. Under these constraints, (\ref{pthree}) successively
reduces to
\bea
\Psurv & & \\ 
(i) & \ra & c_{m12}^2 c_{13}^2 \left\{(1-P_1) c_{12}^2 c_{13}^2 +P_1 s_{12}^2
c_{13}^2 \right\} \\ \nonumber
 &  &+ s_{m12}^2 c_{13}^2 \left\{ P_1 c_{12}^2 c_{13}^2 + (1-P_1)
s_{12}^2 c_{13}^2 \right\} +s_{13}^4 \\ \nonumber
(ii) &\ra &\half + \left( \half - P_1 \right) c_{2\thm_{12}} c_{2\temu}~,
\label{prk}
\eea
which is Parke's formula \cite{parke}.  For large $\nue$ creation densities
($(N_e)^c \gg \res_{12}$), the preceding expression further simplifies to
\beq
\Psurv = s_{12}^2 + P_1 c_{2\temu}~.
\label{limprk}
\eeq

       A number of papers have been devoted to the study of
gravitationally--induced neutrino oscillations
in the two-flavor limit \cite{gasp1,bah1,pant1,mal1,min1}.
All of the results obtained are essentially commensurate with one another,
despite the proposed controversy raised in \cite{bah1}, which assumed a
disagreement in results between \cite{pant1} and \cite{mal1}. There was
an apparent misunderstanding in ref.\
\cite{bah1} of the conclusions in \cite{mal1},
which suggested that the SNU--curve overlap regions for two flavors of
neutrinos were statistically too small to be viable.
That is, according to \cite{mal1} a $\chi^2$ analysis
associates a confidence level of less than $1\%$ for the overlapping
regions, for $2\sigma$ curves.

There are two main regions of overlap which have been found
by all analyses done to date.  These occur for small and large vacuum
mixing angles, respectively.  Figure~2 from \cite{bah1} and Figure~2
from \cite{mal1} can be referenced for comparison.
Ref. \cite{bah1} claims the overlap regions to be:
\begin{itemize}
\item{Nonadiabatic: $\sin^2{2\th} \in [2 \times 10^{-3},10^{-2}]~,~
\df \in [2.7 \times 10^{-14},3.3\times 10^{-14}]$}
\item{Adiabatic: $\sin^2{2\th} \in [0.6,0.9]~,~ \df \in [1.0 \times 10^{-16},
1.5 \times 10^{-15}]$~.}
\end{itemize}

For comparison, \cite{min2} gives the approximate values of the
overlap regions for the two-flavor MSW mechanism as: (1) nonadiabatic:
($\sin^2{2\th} \approx 7\times 10^{-3},\dm \approx 1.75 \times 10^{-3}eV^2$),
and (2) adiabatic: ($\sin^2{2\th} \approx 0.6, \dm \approx 9 \times 10^{-6}
eV^2$).  These masses are roughly equivalent to the above values of $\df$
for neutrinos of energy $\sim 10\:$ MeV, where
we have again  assumed an {\em average} gravitational
potential $|\phi| \sim 5\times 10^{-6}$ for the Sun.

 For large $\temu$ ({\it i.e.} $\sin^2{2\th_{12}} > 0.1$),
the 12--jump term vanishes, {\it i.e.} $P_1 \ra 0$, and $\Psurv$ goes as $s_{12}
^2$,
while for small $\temu$, it is dominated by $P_1$.  From this, we see that
the large $\temu$ solutions (adiabatic approximation) are energy independent,
while the small $\temu$ ones (non--adiabatic) are very much dependent on
the matter effects.

In the adiabatic region, the suppression is energy independent,
and the fluxes of all types of solar neutrinos are equally reduced.
This can be seen from (\ref{limprk}) since the Landau-Zener term $P_{LZ}
\ra 0$ for adiabatic transitions, and we are left with $\Psurv = \sin^2\th$.
Meanwhile, the nonadiabatic mixing region can be found for
small vacuum mixing angles $\th_{12} \sim 10^{-3}$.  
For small $\th_{12}$, the survival probability goes as $P_{LZ}$, which 
is generally large, except near resonance, where it can rapidly 
suppress almost the entire flux.

Using (\ref{prk}), the following results have
been obtained.  First, the $\nue$ suppression probabilities of \cite{bah1}
are accurately reproduced.  The 2--flavor limit was taken by setting
(i) $\res_{13} \gg \nmax$ ($\df_{31} = 10^{-9}$, in this case),
and (ii) $\tetau = 0$.  For all SNU--plots included in this work,
we define the confidence level (CL) as the departure from
the averaged value of the counting rates, as given in Table~\ref{snuexp}.
Hence, a confidence level of 95\% would include all calculated VEP--reduced
rates which fall in the range of $rate \pm 2\sigma$, where $\sigma$ is
the quoted experimental error (see Table~\ref{snuexp}).

Counting rates for neutrinos from source $\alpha$ ({\it e.g.} $\B, \Be$, 
pp, etc...) are obtained via numerical integration of the equation

\beq
R_x^{\alpha} = \int_{0}^{R_{\odot}} \! dr \: r^2 \xi^{\alpha}(r) \int_{E_{min}}^
{E_{max}}
\! dE \: \phi^{\alpha}(E) \sigma_x(E) \Psurv(r,E)~,
\label{count}
\eeq
for neutrinos with energy spectrum $\phi^{\alpha}(E)$ and maximum energy $E_{max
}$,
incident on detector material $x$ with absorption cross--section $\sigma_x(E)$.
The function $\xi^{\alpha}(r)$ represents the fraction of neutrinos of type
$\alpha$ produced at radius $r$.

With cross--sections $\sigma_x(E)$ and neutrino spectra $\phi^{\alpha}(E)$
from \cite{bah3}, we have calculated $2$
and $3\sigma$ overlap regions for a two--flavor limit.  These are shown
in Figs.~\ref{2flava},~\ref{2flavb}.
Figure~\ref{2flava} clearly shows evidence for structure in the regions of the
small and large mixing solutions, differing from the small tail connecting
the two (diagonal line overlap).

In Fig.~\ref{2flavb}, the two overlap regions discussed previously can be seen
amidst the lower statistically--viable areas (e.g. diagonal strip).
A ``new'' region has opened up just above the location of the
small mixing region, but this may simply be a manifestation of the
overestimation mentioned earlier.  Since only experimental errors
are taken into account for this method, this explains the discrepancy
between these plots and the iso--SNU curves of \cite{bah1}.

  The errors in the theoretical fluxes are quite possibly a major
source of the structural difference in the overlap region.  We have
generated plots for both the upper and lower flux limits for the
neutrino sources considered ($\B,\Be,$ and pp).  The $\B$ flux
has an associated error of up to $37\%$, so we should expect this
to greatly determine the allowed regions.  For discussion purposes,
we will use as comparison only the $3\sigma$ level of overlap.

For the lower flux limit case (Fig.~\ref{3l}), the connecting arm is 
greatly expanded, implying that the shift in flux lowers the contours of
one of the experiments more so than the others.  These are most likely
the $\Cl$ contours, since the $\B$ neutrinos are the primary
candidates detected.  The large mixing region tends to drop below
the range indicated by previous studies (see {\it e.g.} \cite{bah1,
shi,zag},...) though, so we can assume
that there is a lower weighting associated to this region of
the flux errors.  The structure of the upper flux limit 
(Fig.~\ref{3u}) is more
like Fig.~\ref{2flavb} than is Fig.~\ref{3l}.  The new region has joined 
with the area of the small mixing solutions, while the large mixing
region has ``fattened''.  Due to the similarities of Figs.~\ref{2flavb}
and \ref{3u}, we can deduce that the statistically--average flux 
contribution to the counting rates tends to favor the higher end of the
flux range, rather than the lower.  A full $\chi^2$ analysis
of the data should reproduce the results cited earlier.

\section{$\Psurv$ Expressions for Specific Resonance Behavior with 
Three Flavors}
\label{resbehav}

\indent Several researchers studying the three--flavor MSW oscillation
mechanism (\cite{msw1,zag,shi})
have concluded that interesting results can be obtained if the 
$\nue$s undergo both $\nue \ra \nm~,~\nue \ra \nt$ resonances, 
{\it i.e.} $\ncr \gg \res_{12,13}$.
Accordingly, $\thm_{12,13} \ra \frac{\pi}{2}$.  
Defining $s_{1i} = \sin{\theta_{1i}}$, $c_{1i} =
\cos{\theta_{1i}}$, and similarly for $s_{m1i}, c_{m1i}$.
the survival probability assumes the simpler form
\beq
\Psurv = P_1 P_2 c_{12}^2 c_{13}^2 + P_2 (1-P_1) s_{12}^2 s_{13}^2
+(1-P_2) s_{13}^2~,
\label{lim1}
\eeq
since all $c_{m1i}$ vanish.

 The matter effects are not gone, due to the
presence of the Landau--Zener jump terms.   
Like the two--flavor case, we examine the small and large
angle effects on (\ref{lim1}), except now we have four cases instead
of two.  The former are generally characterized by an overall $P_i$
dependence, while the latter show energy independence ($P_i \ra 0$).

\subsection{Dependence on $\temu$}
\label{12behavior}

	First, let us consider the small and large $\temu$ cases, to show
how the two--flavor cases change in the full three generation scenario.\\

\noindent \ul{Small $\temu$}:
\beq
\Psurv=s_{13}^2 - P_2 \left\{ s_{13}^2 (1+P_1) - P_1 \right\}~ 
\label{small12}
\eeq
\noindent\ul{Large $\temu$}:
\beq
\Psurv=s_{13}^2 ( 1-P_2 c_{12}^2)~ 
\label{large12}
\eeq
We immediately note a significant departure from the behavior of the
two--flavor mechanism.  In both cases, an explicit $s_{13}^2$ term 
will dominate for an appropriate choice of parameters, and the matter 
effects from the 13--resonance (via $P_2$) are always present.  This 
dependence can have significant impact on the overall flavor conversion 
of solar $\nue$s.  Furthermore, we note
that for small $\tetau$, equation (\ref{small12}) $\ra P_1 P_2$ shows strong
non--adiabatic dependence for both resonances, which can significantly
change the shape of the suppression curve (see Figs.\ref{3prob1},\ref{3prob2}).  
Meanwhile, (\ref{large12}) $\ra 0$ as $s_{13}^2$ becomes small!  Clearly, this
implies an almost complete attenuation of $\nue$s.
This radically different behavior is indicative of the need to further study
the effects of a third flavor.  Note that both of these results are
{\em not} approximations of the two--flavor mechanism, since
the condition $\thm_{13} \ra \frac{\pi}{2}$ is in place.  A recovery
of pure $\nue \ra \nm$ oscillations requires the absence of the
13--resonance, i.e. $\thm_{13} \ra \tetau \ra 0$.
 
\subsection{Dependence on $\tetau$}
\label{13behavior}

	A more relevant discussion of a full three--generation model hinges on
the study of the dependence on the third flavor.  In the previous subsection,
we caught a glimpse of these effects in the small and large mixing regions.
\vskip 0.5 cm

\noindent \ul{Small $\tetau$}:
\beq
\Psurv = c_{12}^2 P_1 P_2~ 
\label{small13}
\eeq
\noindent\ul{Large $\tetau$}:
\beq
\Psurv = s_{13}^2~ 
\label{large13}
\eeq

It can easily be shown that (\ref{small13}) takes on both characteristics
of the previous subsection for small and large $\temu$.  However, perhaps
the most important conclusion to be drawn from the analysis of flavor 
conversion in both 12-- and 13--resonances is (\ref{large13}).
In the three--flavor VEP (and MSW) mechanisms, the presence of a large
$\tetau$ completely dominates the $\nue$ suppression.  This is 
visible in Figs.~\ref{3prob3},\ref{3prob4}.
 
The significance
of this is obvious: it is possible to choose the fraction of surviving
solar $\nue$s by fixing $s_{13}^2$ to the appropriate value.  It is
important to note that this is {\em independent of $\temu$}, and hence
independent of the 12--resonance. 

	We have verified the assertion in \cite{min2} that the very large
$\tetau$ solutions can be ruled out by present data.  Figure~\ref{lrg13}
is clearly indicative that allowed regions can only exist for a very
small range of parameters on the wall near resonance.  The double--resonance
and vacuum oscillations are intuitively ruled out, as in each case,
$\Psurv$ is either too large or too small, respectively.

\subsection{No 12-- or 13--resonances: Vacuum Oscillations}

For values of $\df_{21},\df_{31}$ which exceed $\sim 10^{-12}$, the 
$\nue$s will not interact with matter.  If this situation occurs, then 
the matter--enhanced 
mixing angles $\thm_{12},\thm_{13}$ will take on their corresponding
vacuum counterpart values, {\it i.e.} $\thm_{12} \ra \temu, \thm_{13} \ra
\tetau$.  The probability then reverts to the trivial form
\beq
\Psurv = c_{12}^4 c_{13}^4 + s_{12}^4 c_{13}^4 + s_{13}^4~,
\label{3vac}
\eeq
which is recognizable as (\ref{pthree}), with $V_3^m = V_3$, and 
$P_{LZ} = \id~$.  There are several vacuum oscillation solutions which
are still viable (see \cite{min2} for a brief discussion), 
although they are not considered in this analysis.

\section{Broken Hierarchy: $\df_{21} > \df_{31}$}
\label{hier}

	So far, we have examined the behavior of $\Psurv$ for the two--flavor
limit, as well as for the case of both resonances, assuming a natural 
hierarchy in the violation parameters.  However (apart from 
mathematical and/or aesthetic prejudices) there is no {\it a-priori}
reason to expect there to be such a hierarchy.  In this section we
consider the possibility that $\df_{21} > \df_{31}$, which we refer
to as the broken-hierarchy case.

	In effect, such a break in the natural scheme of things amounts
to a reversal of the matter--enhanced energy--eigenvalues, and hence an
interchange of vacuum and matter violation parameters 
$(f_2,f_3) \ra (F_3,F_2)$ of Eqs.~(\ref{matmass},\ref{abc}).
To show this, we take the limiting form of
these parameters when $f_2 \gg f_3$. 

\subsection{Case 1: No 12--resonance}
\label{bhcaseone}

For this condition to be satisfied, $E\df_{21}$ must be set above the
previously mentioned value.  We assume that $\ncr \gg \res_{31}$, 
in order that 13--resonances still take place.  The matter--enhanced parameters
become dominated by $f_2$, and for convenience we may replace 
$f_2 \ra \df_{21}$ (since we define $f_1 \equiv 0$).
This gives
\bea
\alpha&\sim&\df_{21} (c_{13}^2 c_{12}^2 + s_{13}^2) \\
A&\sim&\df_{21}
\eea
Also, from the definitions of $F_i$ in eq.~(\ref{matmass})
we can show that
\beq
F_3 \sim \df_{21}~,~ F_3 \gg F_2,F_1
\eeq

This implies that
\beq
F_3 \sim \dF_{31} \sim \dF_{32}~,
\eeq
and the additional parameters $\beta,C \ra 0$, while $S \sim 1$.
Therefore, we see that
\bea
s_{m13}^2&=&\frac{F_3^2 - \alpha F_3 + \beta}{\dF_{31} \dF_{32}} \\ \nonumber
 & \sim & \frac{F_3 ( F_3 - \alpha)}{F_3^2} \\ \nonumber
 &  = & \frac{(\df_{21})^2 (1- c_{12}^2 c_{13}^2 - s_{13}^2)}{(\df_{21})^2} \\
\nonumber
\eea
or 
\bea  
s_{m13}^2 & = & s_{12}^2 c_{13}^2~.
\eea
Thus, instead of the naive replacement $s_{m13}^2 \ra s_{12}^2$  that one 
might expect (since we are changing the roles of the eigenvalues),  there 
is a non--trivial dependence on {\em both} vacuum mixing angles.  
If the 13--resonance still takes place, it can similarly be shown that the 
condition $s_{m12}^2 \sim 1$ still holds, {\it i.e.} 
$\thm_{12} \ra \frac{\pi}{2}$.

\subsection{Case 2: No 12-- or 13--resonance}
\label{bhcasetwo}
 
Neither resonance will take place if $\res_{12,13} > \nmax$, thus
the solar $\nue$s will propagate essentially as in vacuum.  However,
due to the ``role--reversal'' of the eigenvalues, the form of the
oscillations will not be the same as (\ref{3vac}).  We have already
seen that the behavior of $s_{m12}^2$ is different than that expected
for $\res_{12} > \nmax$.  In a similar fashion, we will obtain the broken
hierarchy expression for $s_{m13}^2$.

Recall that the matter--enhanced 12--oscillation term is
\bea
s_{m12}^2 & = & \frac{-(F_2^2 - \alpha F_2 + \beta) \dF_{31}}
{\dF_{32} (F_1^2 - \alpha F_1 + \beta) - \dF_{31} (F_2^2 - \alpha F_2
+ \beta)} 
\label{sm12}
\eea

For the scenario in question, the eigenvalues $F_2$ and $F_3$ are
dominated by $f_3$ and $f_2$, respectively.  Accordingly, the numerator 
of (\ref{sm12}) will reduce to
\bea
& & (F_2^2 - \alpha F_2 + \beta) \dF_{31} \nonumber \\
& \ra & f_2 f_3^2 - f_2 f_3 [f_3 c_{13}^2 + f_2 (c_{12}^2 c_{13}^2
+ s_{13}^2)] + f_2^2 f_3 c_{12}^2 c_{13}^2 \nonumber \\
& = & f_2 f_3^2 s_{13}^2 - f_2^2 f_3 s_{13}^2 \nonumber \\
& = & - (f_2^2 f_3 - f_2 f_3^2) s_{13}^2~,
\eea
while the denominator behaves as
\bea
& & \dF_{32} (F_1^2 - \alpha F_1+ \beta) - \dF_{31} (F_2^2 - \alpha F_2
+ \beta) \nonumber \\
& \ra & (f_2 - f_3 ) \beta - f_2 ( f_3^2 - \alpha f_3 + \beta) \nonumber \\
& = & f_2^2 f_3 c_{12}^2 c_{13}^2 - f_2 f_3^2 c_{12}^2 c_{13}^2 - f_2 f_3^2 s_{13}^2
+ f_2^2 f_3 s_{13}^2 \nonumber \\
& = & (f_2^2 f_3 - f_2 f_3^2) (c_{12}^2 c_{13}^2 + s_{13}^2).
\eea
The value of $F_1$ is typically several orders of magnitude smaller than
the other $F_i$, thus we take $F_1/F_i \ra 0$ in the analysis (note that
$F_1 \ra 0$ for full flavor--resonance).  Since we set $f_1=0$,
the values of $f_2,f_3$ and $\df_{21},\df_{31}$ are respectively
identical. Thus, (\ref{sm12}) reduces to a simple expression.  For vacuum 
oscillations in a broken hierarchy scheme, the matter--enhanced mixing 
angles assume the modified forms
\bea
&s_{m12}^2 = \frac{s_{13}^2}{s_{13}^2 + c_{12}^2 c_{13}^2}~ 
\label{bhvaca}\\
&s_{m13}^2 = s_{12}^2 c_{13}^2~.
\label{bhvacb}
\eea

In general, the resulting suppression will be smaller for a broken hierarchy
than for a natural one (as shown in Figs.~\ref{3prob3},~\ref{3prob4}). 
For small $\tetau$, we see that (\ref{bhvaca}) approaches 0, while
(\ref{bhvacb}) approaches $s_{12}^2$.  This is consistent with the decoupling
of the third neutrino, except here the decoupled flavor is the $\nm$.
If both $\temu$ and $\tetau$ are small,
the natural and broken hierarchical values of $\Psurv$ are almost identical.
For example, in the case of Fig.~\ref{3prob1}, the difference has been
calculated as less than 0.001\% for various violation parameter values (with
$\df_{21} > \df_{31}$).
 
This implies that a broken hierarchy in the
vacuum oscillation range would be almost indistinguishable from the natural
case.  
 
For larger angles, the fact that the flux reduction is greater
than for the natural case can possibly be used to our advantage: a specific
choice of mixing angles can lead to $\Psurv$ which is too large to account
for the observed fluxes.  Conversely, the broken hierarchy vacuum oscillation
$\Psurv$ is smaller, and can conceivably fit the data in situations
when the former cannot.

\section{$\B, \Be$, and pp  Suppression}
\label{nusup}

	In the two--flavor case, we saw that the suppression pits for neutrinos
of varying energies differed in size from each other (see \cite{bah1}).
In the following section, we will briefly examine these variations in the
three--flavor model.  For simplicity, we compare the $\B, \Be$, and pp
neutrinos only in the large $\temu$ case.  The curves for small $\temu$ show
similar behavior.  Note that all axes of the form $2E\df_{i1}$ will be 
expressed in units of $10^{18}\:$MeV$^{-1}$.

In Fig.~\ref{lrgsm}, we see how the suppression radically varies over several
orders of magnitude in $2E\df_{31}$.  We first consider the case
of small $\tetau$, to show how the two--flavor results are affected by the
addition of the $\nt$ resonance.  For relatively small values,
we see the behavior indicated in Section~\ref{13behavior}, namely that small values
of $\tetau$ can lead to almost complete attenuation of the flux.  The width
of the suppression pit is almost the same as for the two--flavor scenario,
which is approximated by the upper set of curves 
($2E\df_{31} = 2\times10^{-6}$).
 
However, several differences to note include the following: while each
type of neutrino is equally suppressed along the upper pit--bottom, there
is a greater amount of suppression for the pp and $\Be$ neutrinos in the 
lower curve. 
 
In fact, for the lower curve, the suppressions are too great to match
the experimental values cited by the solar neutrino experiments.
Also, for a broken hierarchy with one resonant flavor (visible in the lower curve 
of Fig.~\ref{lrgsm} for $2E\df_{21} > 10^{7}$),
we note that the pp neutrinos are suppressed {\em less} than the $\Be$s 
and $\B$s. 
 
This should be contrasted with the regular two--flavor case, 
where in general the pp pits always end before those of the $\B$s.  
The kink behaviors visible in the upper curve  of 
Fig.~\ref{lrgsm} are a result of crossing
the hierarchy--boundary ({i.e.} the surface $2E\df_{21} = 2E\df_{31}$).

In contrast, Fig.~\ref{lrglrg} shows the same curves for large $\tetau$.  
The dominance of $s_{13}^2$ is clearly visible for the 
curve $2E\df_{31} = 2\times 10^{4}$. 
 
Also visible is the inversion of energy suppression for the
case of the broken hierarchy.  In the case of only $\nue \ra \nt$ resonance,
the pp neutrinos are suppressed less than the $\B$ neutrinos.
The full effect of the third flavor is visible in the
$2E\df_{31} = 2\times 10^6$ curve .  This represents the $\nt$
above resonance, but with large mixing angle $\tetau$, and should be compared
with the two--flavor case from \cite{bah1}.  The variation in suppression
of differing energies is also very striking: the pp neutrinos are suppressed
much more than are the $\B$s in this case, such that the flat pit bottoms
do not even overlap.  The average energies of the $\Be$ and pp neutrinos are
an order of magnitude less than those of the $\B$s.  Hence, when the $\B$ 
neutrinos have energies $2E\df_{31} > 10^{-12}$, the pp and $\Be$ will not.  
Thus, there can be a double resonance behavior for certain types of neutrinos,
while others have only one.

This yields interesting possibilities if experiment
dictates the various fluxes are suppressed by different amounts.  However,
since it is generally the case that the $\B$ flux is attenuated to a higher
degree than is the pp flux, this behavior probably cannot account for the
flux deficit.  The broken hierarchy single--resonance cases discussed earlier
can provide viable solutions if data requires varying degrees of reduction
for each flux, since the $\B$ are more suppressed than the
pp neutrinos are.

In the regular two--flavor case, overlap regions requiring
differently suppressed fluxes can only occur on the pit walls, implying
only a small allowed area of parameter space.  Adjacent flat--bottom $\Psurv$
curves of differing values can provide a much larger acceptable parameter space
region to match experiment.

	For the case of extremely large $\tetau$ mentioned in Section~\ref{13behavior},
Fig.~\ref{big13} shows curves for $2E\df_{31} = 2\times 10^4$ and $2\times 10^6$.
For both sets of curves, the neutrinos of lower energies are always 
suppressed to a greater
extent than those of higher energies (expect for the $2E\df_{31}$ curve, which 
goes as $s_{13}^2 = 0.8$, as with large $\tetau$ behavior, before the resonance boundary).  
For the leftmost curves, the pp flux can be attenuated by almost a factor of
2 more than the $\B$ neutrinos in places along the pit wall, with the $\Be$ suppression
somewhere in between.  Depending on the input values of the fluxes, this can
perhaps provide a viable solution.  The $2E\df_{31} = 2\times 10^6$ curve 
shows much more constraining behavior, though,
with an extremely steep pit wall (partially due to hierarchical inversion) 
over a short range of $2E\df_{21}$ values , much akin to the pit wall
of the non--adiabatic 2--flavor solution.  It seems rather unlikely that a
statistically significant overlap region of parameter
space can open for these energies, since the method of calculation used in
this paper tends to overestimate the allowed regions.  

\section{Parameter Space Analysis for Three--Flavor Oscillations}
\label{threesnu}

	In Section~\ref{nusup}, we showed how the addition of a third
flavor affects the suppression of $\nue$ fluxes for varying sources
of neutrinos.  Different types of neutrino fluxes ($\B, \Be$, pp) are 
suppressed to 
differing degrees, depending on specific conditions such as
double--resonance and single--resonance behavior, vacuum oscillations,
broken hierarchies, and so forth.  In this section we consider how
all of these effects come together by studying the parameter
space overlap regions for the experimental values cited in
Table~\ref{snuexp}.  

	We shall focus on the regions of overlap as calculated
by eq.~(\ref{count}).  We present plots for 2 and 3$\sigma$ C.L.s, as defined
in Section~\ref{twosnu}, for the limiting conditions discussed in 
Section~\ref{resbehav}
{\it i.e.} large and small $\temu$ and $\tetau$ values.  We find that,
particularly for the large $\tetau$ solution, large areas of parameter
space become viable overlap regions at the $3\sigma$ level. 

	As previously mentioned, one of the nicest ways to visualize
the effect of the various oscillation models on solar neutrino
rates is to look at the regions of parameter space which allow such
solutions.  In Section~\ref{twosnu}, we saw the limiting case of two flavors,
and how this compared with other two--generation studies.  Again, 
before reviewing the following results, 
it should be stressed that the allowed regions of
parameter space discovered in this Section have been done so by 
straight numerical overlap, and hence can overestimate certain regions
(again, see \cite{hata} for a detailed discussion
on the different techniques for error treatment).  Since the parameter
space in question is actually four dimensional ($\temu,\tetau,
\df_{21},\df_{31}$), we present the regions as ``slices'' of fixed
($\tetau,\df_{31}$).  This is a natural choice, as it facilitates 
comparison with the allowed two--flavor SNU overlaps.  We overlay
the results with the two--generation allowed boundaries (see the
referenced figures in \cite{bah1,mal1}),
to show how the new degrees of freedom can
introduced regions outside of these areas.

\subsection{Small $\tetau$}

	So long as $\res_{13} > \nmax$, sufficiently small values
of $\tetau$ will recover the two--flavor model.  However, as noted
in Section~\ref{resbehav}, if $\df_{31} < 10^{-12}$, then $s_{m13}^2 \ra
1$, and $\Psurv$ will take on a different form than
for a single resonance.  Unfortunately, (\ref{large12}) shows that 
the neutrino fluxes can almost be extinguished for the 
double--resonance region, where both $\df_{21}, \df_{31} <
10^{-12}$. If this
is the case, then an overlap in parameter space for these values
would most likely only appear at the $3\sigma$ C.L., as defined in
this work.   

	Figure~\ref{3f13s001} attests to the low survival rate,
showing only overlap regions for a broken hierarchy in the small
$\temu$ range.  This region is mostly due to the nonadiabatic
contributions from the Landau--Zener Jump terms (see eq.~(\ref{small13})).
At the $2\sigma$ C.L., the overlap regions completely vanish, so
the plots are not included.  Looking at the upper--flux limits
for each neutrino source, though, we see a slightly different 
picture.  Figure~\ref{u3f13s001} shows the $3\sigma$ overlap regions
for the same choice of parameters employed in Fig.~\ref{3f13s001}, using
only the upper bounds for each experimentally--observed rate.  We
see that a much larger region has opened up for this choice of
fluxes, which implies that a small $\tetau$ solution is very sensitive
to the value of the flux.  The lower flux range plot is very similar
to Fig.~\ref{3f13s001}. As the third flavor begins to decouple, 
{\it i.e.} $\df_{31} \sim 10^{-12}$, the overlap regions for 
$s_{13}^2 = 10^{-3}$ vanish all together.

However, as we move down an order of magnitude, regions begin to open
up once more. Figs.~\ref{3f13s0001} and \ref{3f12s0001} show these
regions for $s_{13}^2 = 10^{-4}$.  Note that the region in Fig.~\ref{3f12s0001}
begins to take the form of the two--flavor region from Fig.~\ref{2flavb}.

\subsection{Large $\tetau$}
\label{lrg13snu}

We begin to see new regions open up for values of $\tetau$
which were not allowed for double--resonance.  For example, large 
$s_{13}^2$ solutions begin to become more abundant.
Figs.~\ref{3f12s2},~\ref{3f12s3},~\ref{3f12s4} show how the regions 
evolve with increasing 13--mixing angle.  It should be noted that
such behavior is {\em not} apparent for the previously discussed
case of $\df_{31} = 10^{-13}$.  The only viable region in that
case is $s_{13}^2 = 0.4$.  Lower values of  $s_{13}^2$ show previously
unallowed  vacuum oscillation regions for a broken hierarchy, 
much like Fig.~\ref{3f13s001}.

	This is due to the strong 
dependence on $s_{13}^2$ from the probabilities
of Section~\ref{large13}.  In fact, for a double--resonance (where the
survival probability goes as $\Psurv = s_{13}^2$) we should expect to
see most of parameter space become viable.  Indeed, Fig.~\ref{3f13s4} shows
exactly this situation.  The two--flavor boundaries are clearly ignored in this
case, as the 13--resonance dominates the suppression.  The problem 
with such a solution is that
despite the excellent mobility it gives one in parameter space, the
constraint imposed by the limiting form of $\Psurv$ severely impedes
much leeway from set counting rates.

Fig.~\ref{2f13s4} testifies to this fact.  Essentially, this is a 
result of the counting
rates being ``fixed'' to a value of $R_{VEP} \sim R_{SSM} \times s_{13}^2$.  
If $R_{VEP}$ is in the range $R_{expt}\pm 3\sigma_{expt}$, but not
in the range  $R_{expt}\pm 2\sigma_{expt}$, then Figs.~\ref{3f13s4}
and ~\ref{2f13s4} will be the result for the overlap regions.

	If we examine the $3\sigma$ upper--flux limit of the large 
mixing region,
then we see even more interesting behavior.  Fig.~\ref{u3f13s4} shows
large bands which do not constitute viable regions of parameter space,
with a central accepted area,  and the regions below the 2--flavor 
adiabaticity boundary disappear.  The sensitivity of the allowed 
overlaps becomes very apparent for large $s_{13}^2$ in this  case.

\subsection{SNU--Region Comparison: VEP v.s.\ MSW}
\label{snucomp}

	We have seen how the SNU regions are commensurate with those calculated
for the two--flavor VEP mechanism (see Section~\ref{twosnu}), and have also discussed
the commonalities between two--flavor VEP and MSW regions.  Here
we compare the full three-generation VEP mechanism 
to previous findings from three--flavor MSW studies.  However 
most studies of the MSW three-flavour case
concentrate on a combined analysis of data from all types
of neutrino experiments.  We shall reference the plots in \cite{shi},
as these are exclusively fits of solar neutrino counting rates.

	Ref. \cite{shi} presents the MSW overlap regions for the two--flavor
limit, as well as small and large $\tetau$ solutions, at the $1,2$, and
$3\sigma$ C.L.s.  Although the definition of the confidence level differs 
from that used here, it is still possible to compare the works based simply 
on the behavior of the regions for the different $\sigma$ limits.  In 
particular, \cite{shi} shows that the two--flavor limit overlap zones 
occur exclusively on the diagonal and vertical regions of the plots.  
This is due to the
wide variation of $\nue$ survival probability behavior for the 
small and large 12--parameters ($\df_{21}$ or $\dm_{21}$ and $\temu$).
The horizontal solution is ruled out in this analysis as it implies
a suppression of mostly high energy neutrinos, which contradicts the
observed KII data (sensitive {\em only} to high energy neutrinos).
The vertical solution represents energy--independent suppression
(large--angle $\Psurv$ behavior, see Section~\ref{twosnu}).

	Meanwhile, for non--zero $\tetau$, the results are also very
similar.  The $1\sigma$ overlap regions are thin, and for the most
part, do not exist on the horizontal or vertical portions of the
SNU curves.  As the analysis proceeds to the $3\sigma$ level, we
see regions expanding and new ones opening up all over the parameter
space in question.  The work in \cite{shi} shows extremely large
allowed regions for large $\tetau$ at the $3\sigma$ level, which is  similar
to the results of Figs.~\ref{3f12s4},~\ref{3f13s4}.

\section{Discussion}
\label{ch6}

We have examined the viability of extending the VEP oscillation mechanism
from two neutrino flavors to the realistic three--generation model as 
a possible resolution to the SNP.  We have
studied the behavior of the $\nue$ survival probability $\Psurv$ in various
limiting cases, including a recovery of the two--flavor mechanisms.   
The concept
of a broken hierarchy was compared to the natural one, to see how the dynamics
could be used to one's advantage as a successful resolution of the SNP. 
A resulting
examination of the SNU overlap regions in parameter space for the four
cited experiments ensued.

	So what conclusions can be drawn for the three--flavor VEP model?
Does it expand or create new regions in parameter space?   Certainly, the
probability analysis of section ~4 tends to point in favor of expansion; 
the dominance
of large $\tetau$ can yield enormous viable regions of parameter space.  This
was demonstrated in the SNU plots of Section~\ref{snucomp}.  However, 
due to the rigid constraints on the $\nue$--suppression 
(see section~\ref{lrg13snu}), these
regions hold only for large enough $\tetau$, and so generally disappear for
lower confidence levels.  A broken hierarchy opens the possibility of small
$\temu$ oscillations, particularly if these occur in a single--resonance zone
for $\nue \ra \nt$ transitions only.

	When compared to the overlap regions calculated in \cite{shi}, we
find that the results obtained in this paper are similar to those
for an MSW analysis.  In the limiting two--flavor case, they 
find that the only allowed regions are on the diagonal and vertical
parts of the SNU curves, which is consistent with the figures of Section~\ref{twosnu}.
This is due to the varied suppression in the fluxes (as discussed in
Section~\ref{resbehav}).  The diagonal is highly dependent on the nonadiabatic jump $P_1$,
while the vertical is representative of equal (energy independent) suppression
for all fluxes ({\it i.e.} $\: \sim s_{12}^2$).  At the quoted $3\sigma$ 
experimental levels of Table~\ref{snuexp}, these are both allowed.

	For both two and three generation models,  low
confidence level overlap regions do not exist for either the horizontal
or vertical areas ({\it i.e.} only on the vertical, or non--adiabatic, region).  
For higher
C.L.s, they show extremely large and varied regions of overlap opening
up.  Up to method of calculation, these results are commensurate with
the SNU plots presented in Sections~\ref{twosnu} and \ref{threesnu}.  
In particular, the $3\sigma\;$ large $\tetau$  SNU plots of \cite{shi} 
are very similar in structure to those of Section~\ref{threesnu}.  
We have verified the assertion in \cite{min2} that
a very large value of $s_{13}^2$ is effectively ruled out by the solar
neutrino data.  Since the work in \cite{min2} is based largely on a
combined analysis of solar neutrino, oscillation (LSND), and reactor 
(Bugey) experiment data, no further comparisons can be made at this time.

	We point out that, while VEP and MSW yield similar SNU plots,
the differing energy dependence in eq.~(\ref{ecomp})
should manifest
itself in one way or another.  The most direct way of detecting such
a variation would be through observation of the neutrino 
spectrum.  The authors have performed such a comparison
\cite{jrm2}, and have show that for certain values of the oscillation
parameters, the $\nue$--flux can be suppressed in radically different
ways for each model (at a fixed counting rate).

How does the VEP analysis hold up in light of the reported results 
of non--solar neutrino experiments?  It has been 
suggested \cite{ut1} that a two--flavor model is insufficient to explain 
solar, atmospheric, and reactor neutrino data all at once (unless the
potential of the supercluster dominates \cite{pant2}); furthermore
a degenerate-mass VEP mechanism could
be ruled out by an improvement of the recent LSND experiments \cite{ut1}.

Other possible realizations of the VEP mechanism await more detailed study. 
For example, the effects of $\nue$--regeneration as the solar neutrinos 
pass through the Earth can provide yet another test for both the
VEP and MSW mechanisms. Presently the detected asymmetry between day (D) 
and night (N) rates is $\frac{N-D}{N+D} =0.07 \pm 0.08$ \cite{dayn} 
and so such observations need a considerable amount of refinement before
such a test would be viable. Another viewpoint on the VEP
mechanism was recently expressed in ref. \cite{pant2}, in which  
the dominant potential that induces flavor oscillation 
was assumed to be that
of the Great Attractor (or local supercluster), which is constant,
and larger than the maximum solar potential by almost
an order of magnitude ($10^{-5}$).  A similar recent study of high
energy neutrinos from distant AGN \cite{newmin} supposes that this 
potential will be the one neutrinos couple to as they propagate through
the intergalactic medium.  An investigation of the effects of
various potentials (constant and varying) on VEP oscillations is currently
under way \cite{jrm3}.  

Clearly, the notion of testing the Equivalence
Principle via neutrino flavor conversions is still an active
and exciting field, promising many new and interesting results.
With the advent of such detectors as Superkamiokande \cite{superk}
and SNO \cite{sno}, 
able to measure at the least the {\em entire}
(flavor--independent) solar neutrino specrtum, the SNP may be on
its way to becoming the SNE, or Solar Neutrino Effect.  

\vskip .25 cm

\noindent
{\bf Acknowledgements}
 
This work was supported in part by the Natural Sciences and Engineering
Research Council of Canada.

\pagebreak


\begin{table}
\begin{center}
{\begin{tabular}{||c|c|c|c||}\hline
Expt. & $R_{\nue}$ & $\Delta R$ & Units \\ \hline\hline
Homestake & 2.55 &  0.25 & SNU \\ \hline
SAGE & 73  & 19.3  & SNU \\ \hline
GALLEX & 79 & 11.7 & SNU \\ \hline
KII & 0.51  & 0.07 & $\Phi^{SSM}_{\B}$\\ \hline \hline
\end{tabular}}
\end{center}
\caption{Recent reported rates from various solar neutrino 
detector experiments.  Errors $(\Delta R)$ are quoted to 
$1\sigma$--uncertainty.}
\label{snuexp}
\end{table}

\begin{table}
\begin{center}
{\begin{tabular}{||c|c|c|c|c|c||}\hline
${\cal M}$ & \multicolumn{4}{c|}{$R^{SSM}_{\nue}$} & Total \\ \cline{2-5}
     & $\B$  & $\Be$  & pp & Other & \\ \hline\hline
$\Cl$ & 6.1 & 1.1 & 0.0 & 0.9 & $8.0 \pm 1.0$ SNU\\ \hline
$\Ga$ & 14.0 & 34.3 & 70.8 & 11.9 & $132^{+20}_{-17}$ SNU \\ \hline
$\nue$-$e$ & 5.8& 0.0 & 0.0 & 0.0 & $5.8 \pm 2.1 \times 10^{6}$cm$^{-2}$s$^{-1}$ \\ \hline\hline
\end{tabular}}
\end{center}
\caption{Theoretical counting rates for material--${\cal M}$--based detectors.}
\label{snussm}
\end{table}

\pagebreak

\begin{figure}[h]
\begin{center}
\leavevmode
\epsfysize=300pt
\epsfbox[-80 80 740 710] {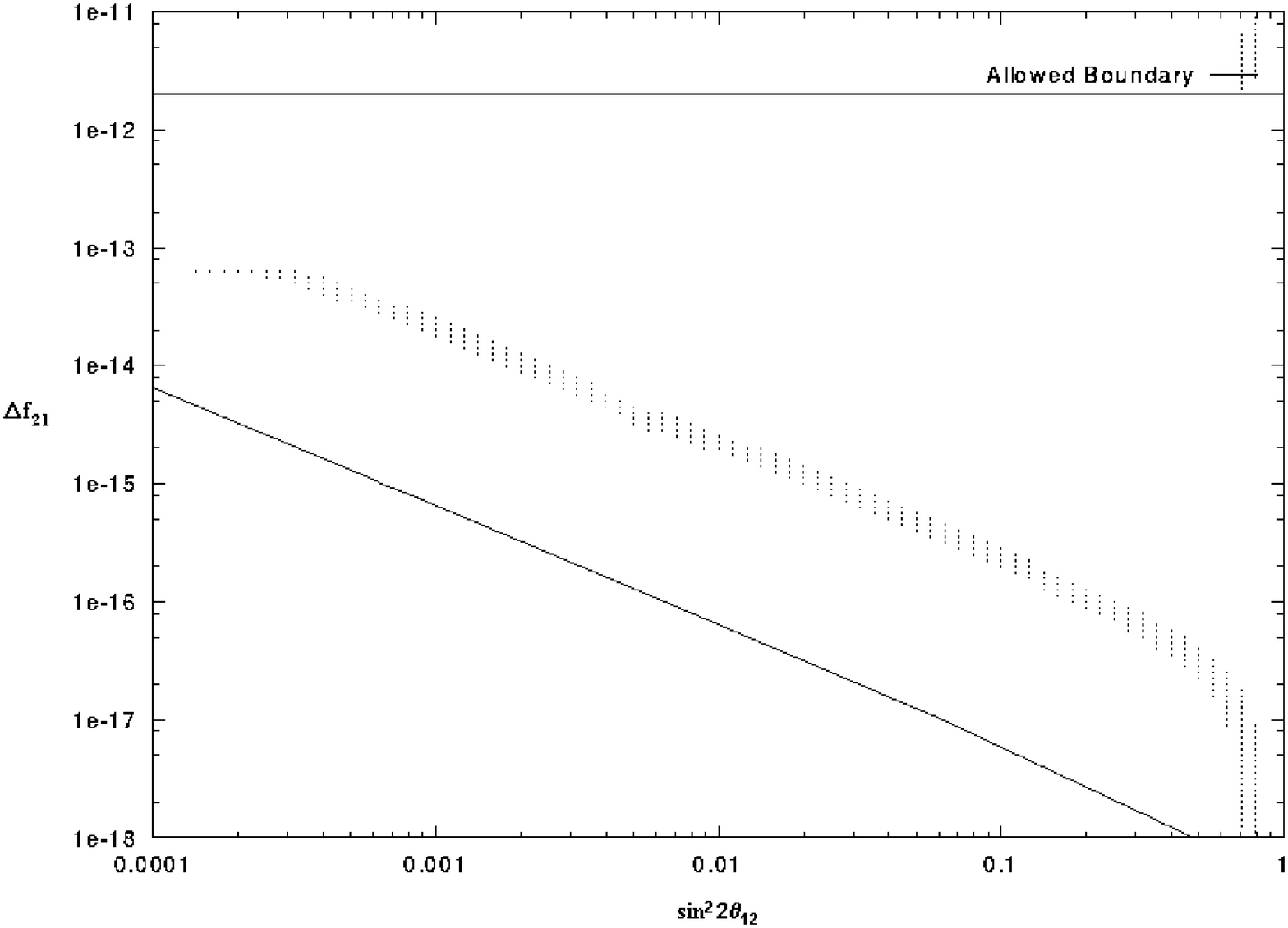}
\end{center}
\caption{2--flavor limit, $2\sigma$ range.}
\label{2flava}
\end{figure}

\begin{figure}[h]
\begin{center}
\leavevmode
\epsfysize=300pt
\epsfbox[-80 80 740 710] {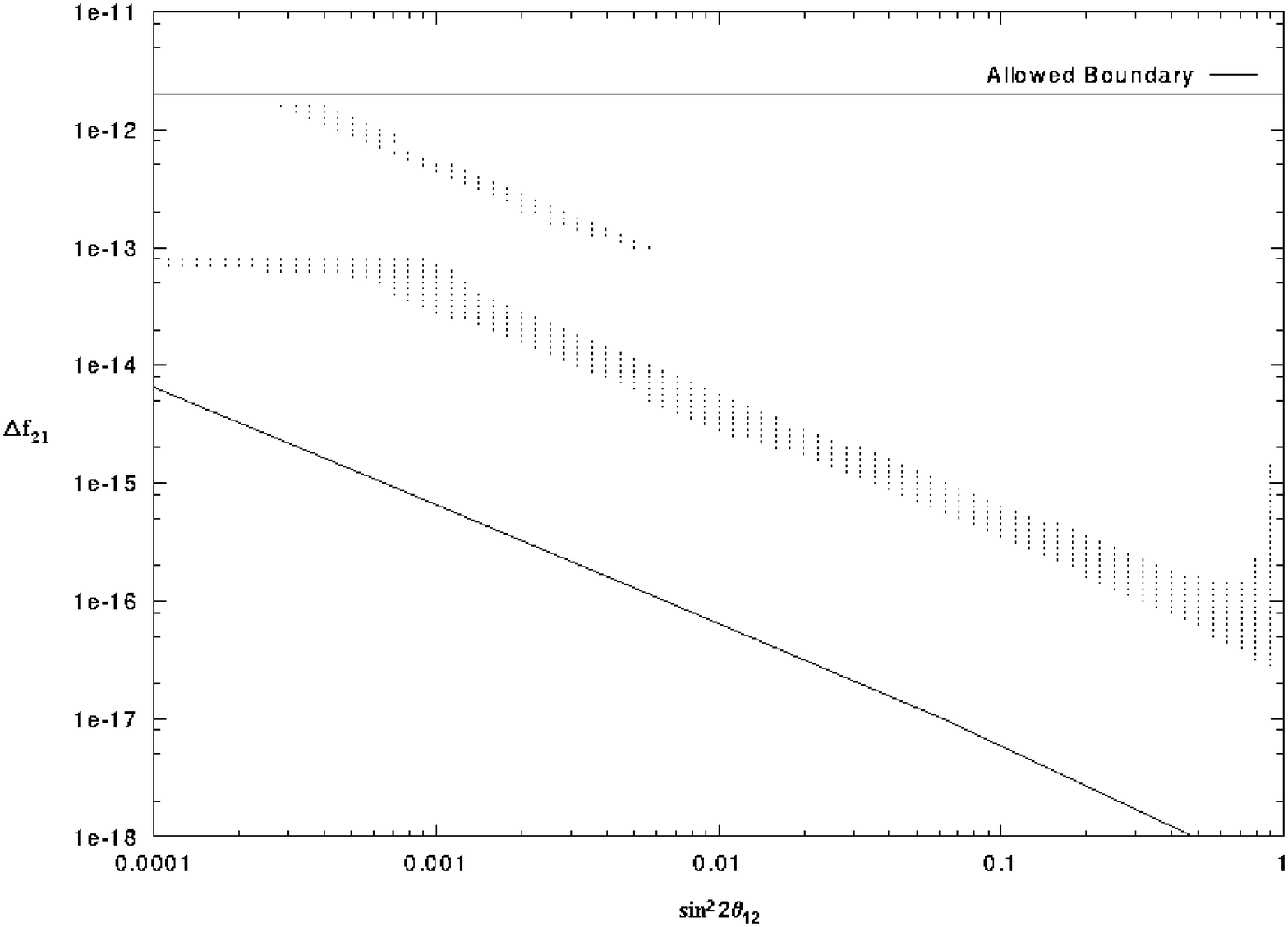}
\end{center}
\caption{2--flavor limit, $3\sigma$ range.}
\label{2flavb}
\end{figure}

\begin{figure}[h]
\begin{center}
\leavevmode
\epsfysize=300pt
\epsfbox[-80 80 740 710] {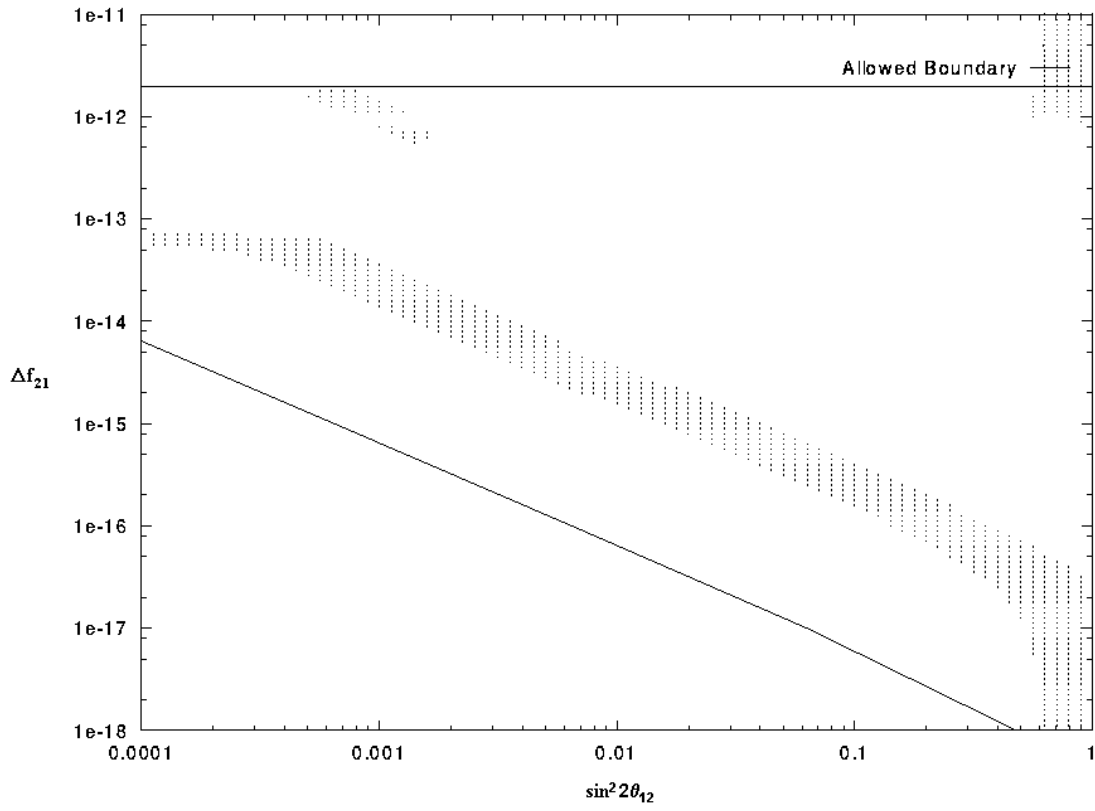}
\end{center}
\caption{Lower flux limit, 2--flavors, $3\sigma$ range.}
\label{3l}
\end{figure}

\begin{figure}[h]
\begin{center}
\leavevmode
\epsfysize=300pt
\epsfbox[-80 80 740 710] {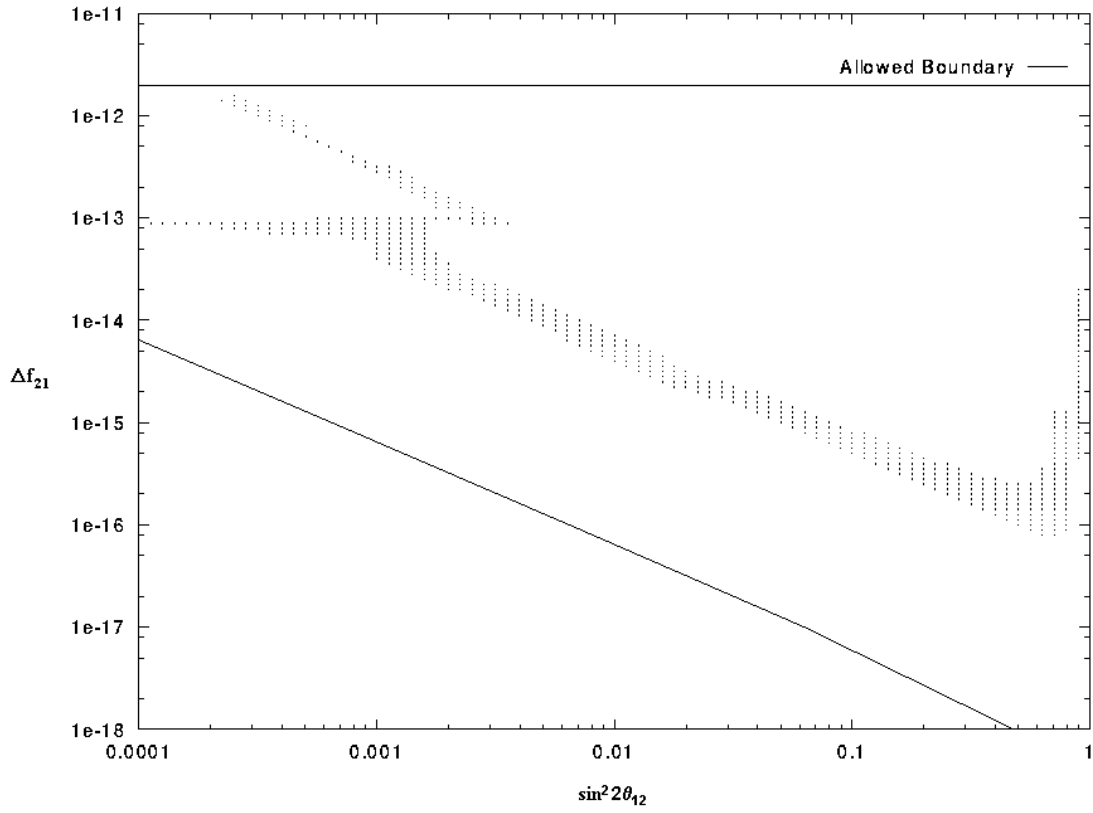}
\end{center}
\caption{Upper flux limit, 2--flavors, $3\sigma$ range.}
\label{3u}
\end{figure}

\begin{figure}[h]
\begin{center}
\leavevmode
\epsfysize=290pt
\epsfbox[-230 -23 825 784] {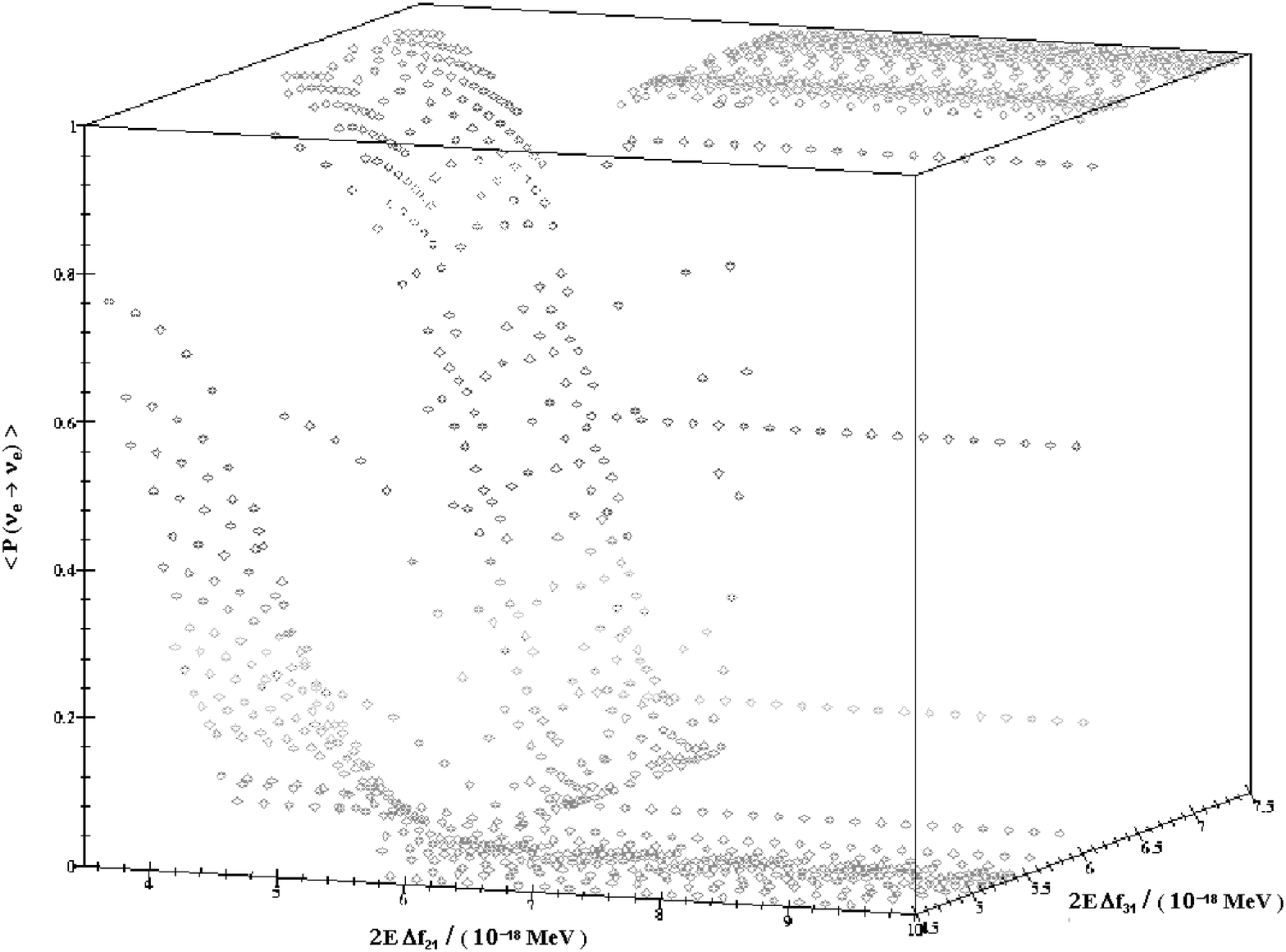}
\end{center}
\caption{$\sin^2 2\temu = 5\times 10^{-3} ; \sin^2 \tetau =  10^{-3}$}
\label{3prob1}
\end{figure}

\begin{figure}[h]
\begin{center}
\leavevmode
\epsfysize=290pt
\epsfbox[-230 -23 825 784] {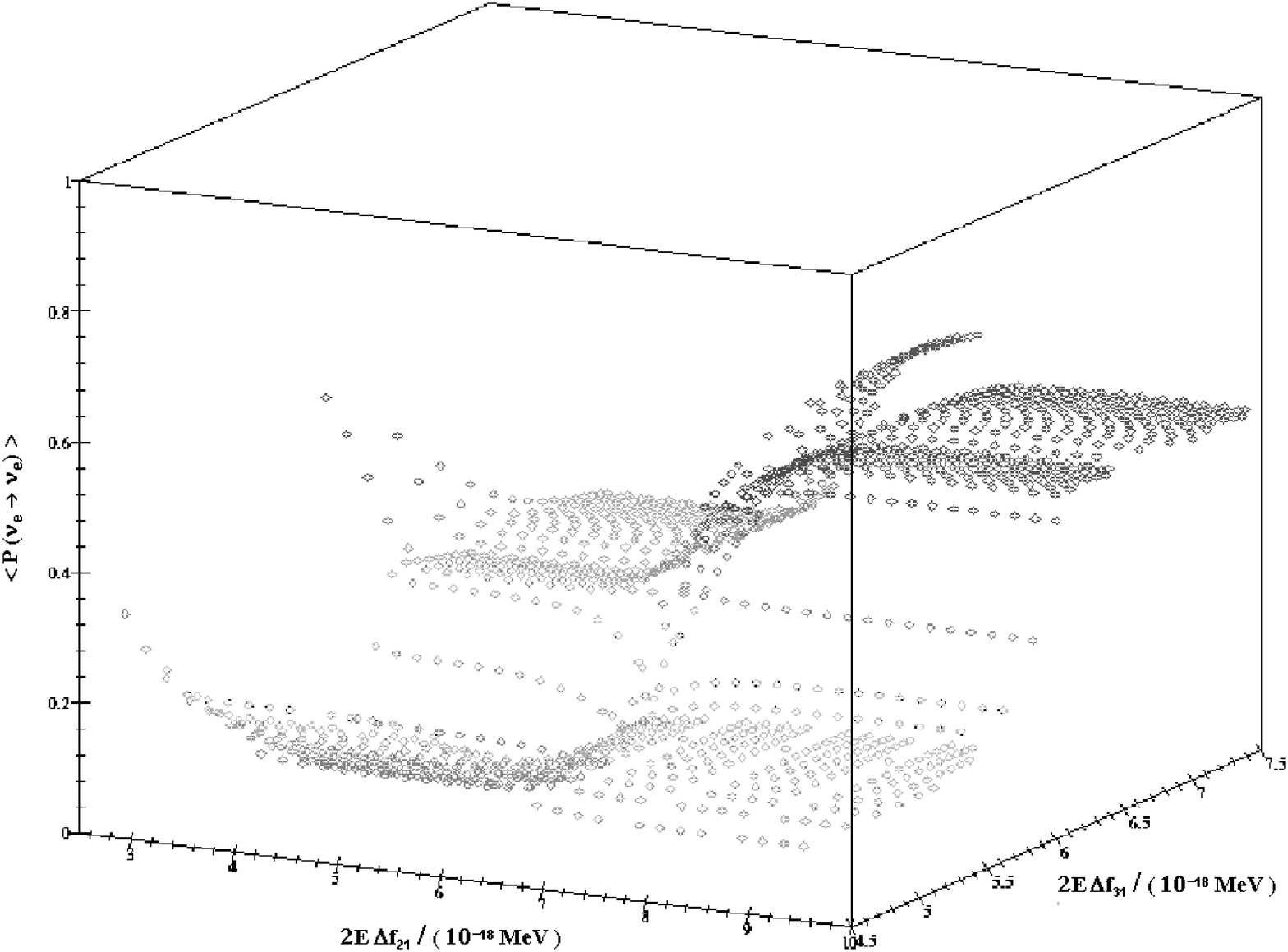}
\end{center}
\caption{$\sin^2 2\temu = 0.8 ; \sin^2 \tetau = 10^{-3}$}
\label{3prob2}
\end{figure}

\begin{figure}[h]
\begin{center}
\leavevmode
\epsfysize=290pt
\epsfbox[-230 -23 825 784] {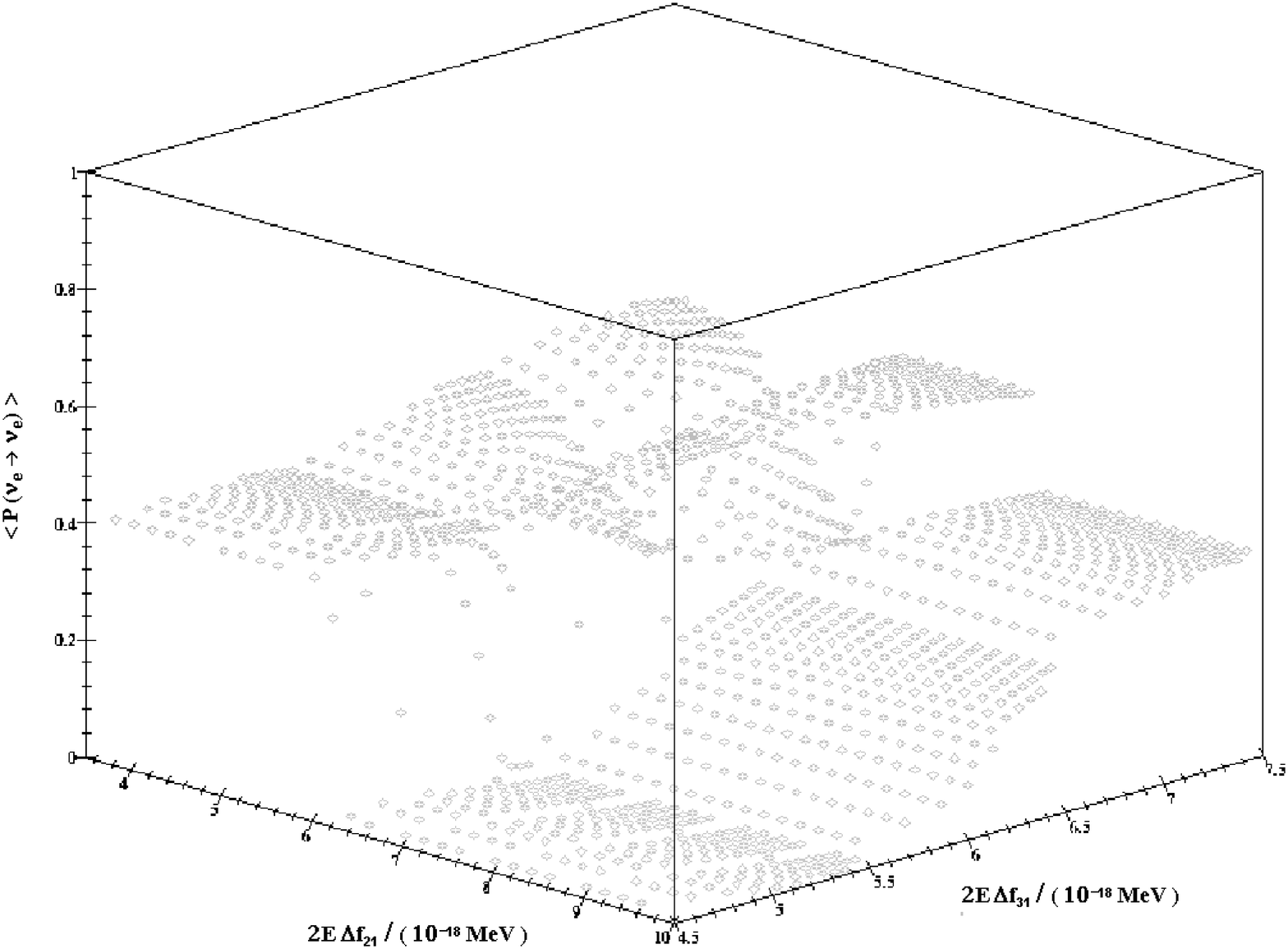}
\end{center}
\caption{$\sin^2 2\temu = 5\times 10^{-3} ; \sin^2 \tetau = 0.4$}
\label{3prob3}
\end{figure}

\begin{figure}[h]
\begin{center}
\leavevmode
\epsfysize=290pt
\epsfbox[-230 -23 825 784] {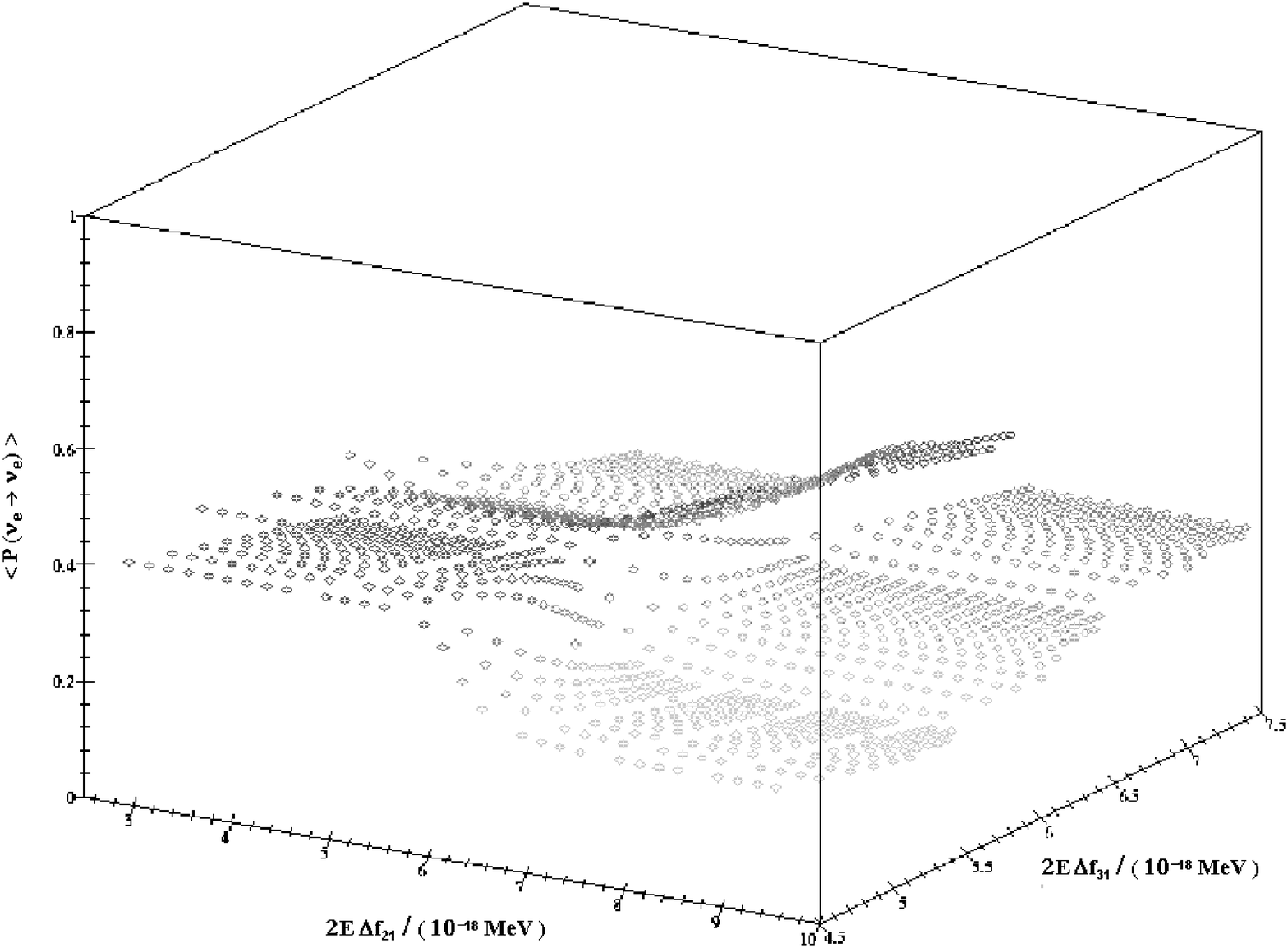}
\end{center}
\caption{$\sin^2 2\temu = 0.8 ; \sin^2 \tetau = 0.4$}
\label{3prob4}
\end{figure}

\begin{figure}[h]
\begin{center}
\leavevmode
\epsfysize=290pt
\epsfbox[-230 -23 825 784] {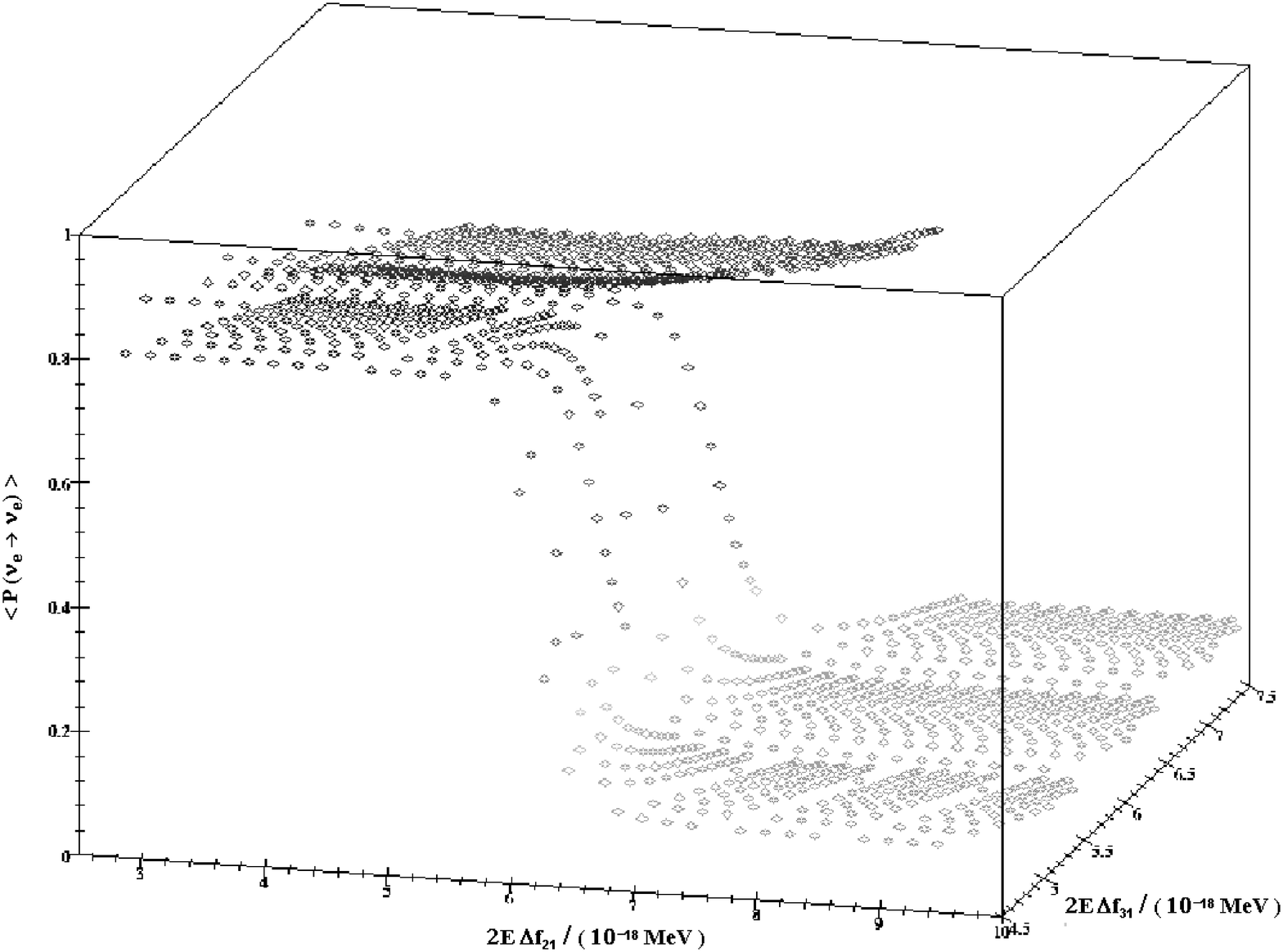}
\end{center}
\caption{$\sin^2 2\temu = 0.8~ ;~ \sin^2 \tetau =  0.8$}
\label{lrg13}
\end{figure}

\begin{figure}[h]
\begin{center}
\leavevmode
\epsfysize=290pt
\epsfbox[-145 73 753 725] {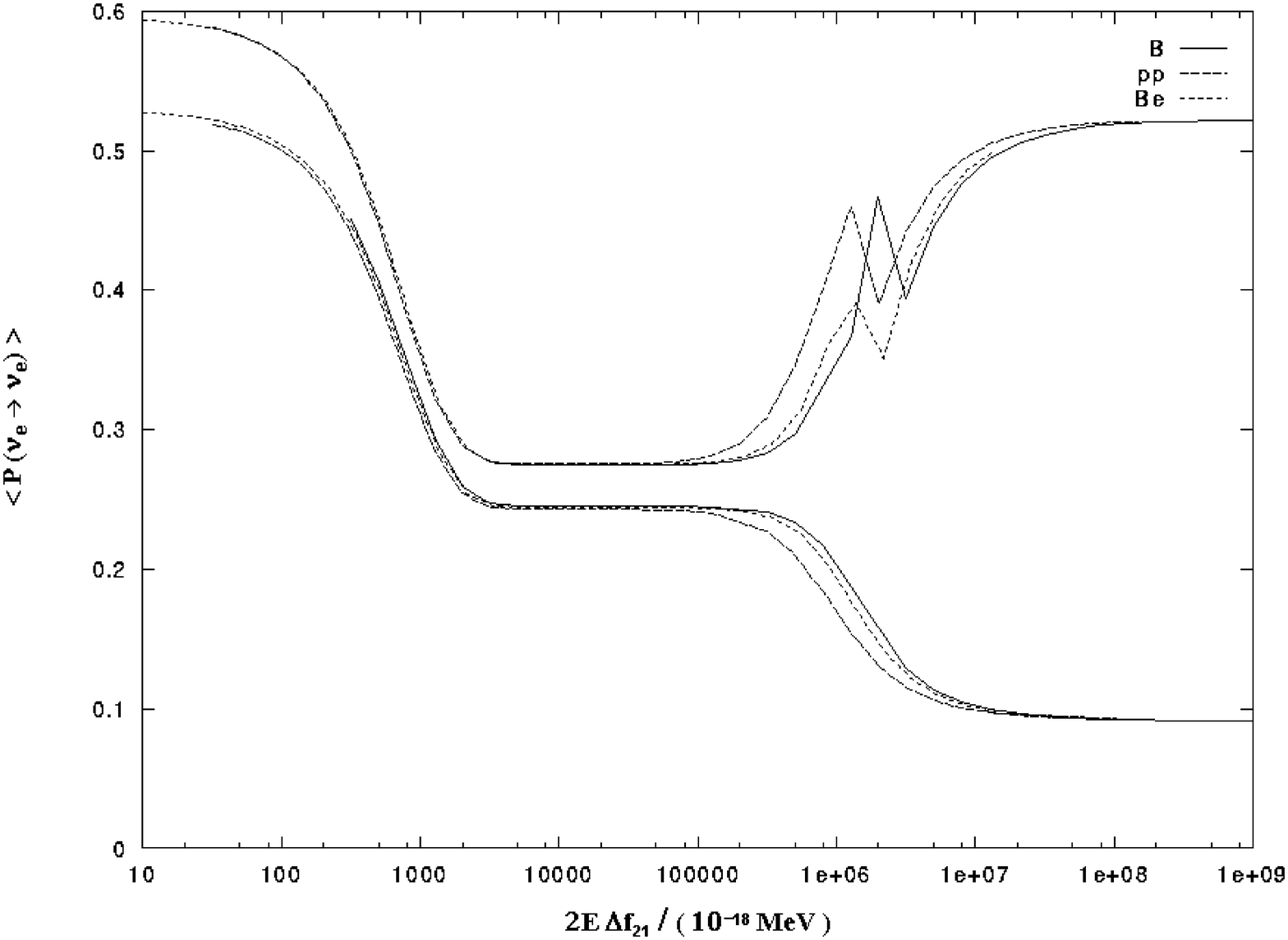}
\end{center}
\caption{$\Psurv$~for~$\B$,$~\Be$,~and~pp neutrinos, fixed values of  $2E\df_{31}/(10^{-18}\:$MeV) = $2\times
10^4$ (lower), and $2\times 10^6$ (upper).  
$s_{2\th_{12}}=0.8~;~s_{13}^2 = 10^{-3}$}
\label{lrgsm}
\end{figure}

\begin{figure}[h]
\begin{center}
\leavevmode
\epsfysize=290pt
\epsfbox[-145 73 753 725] {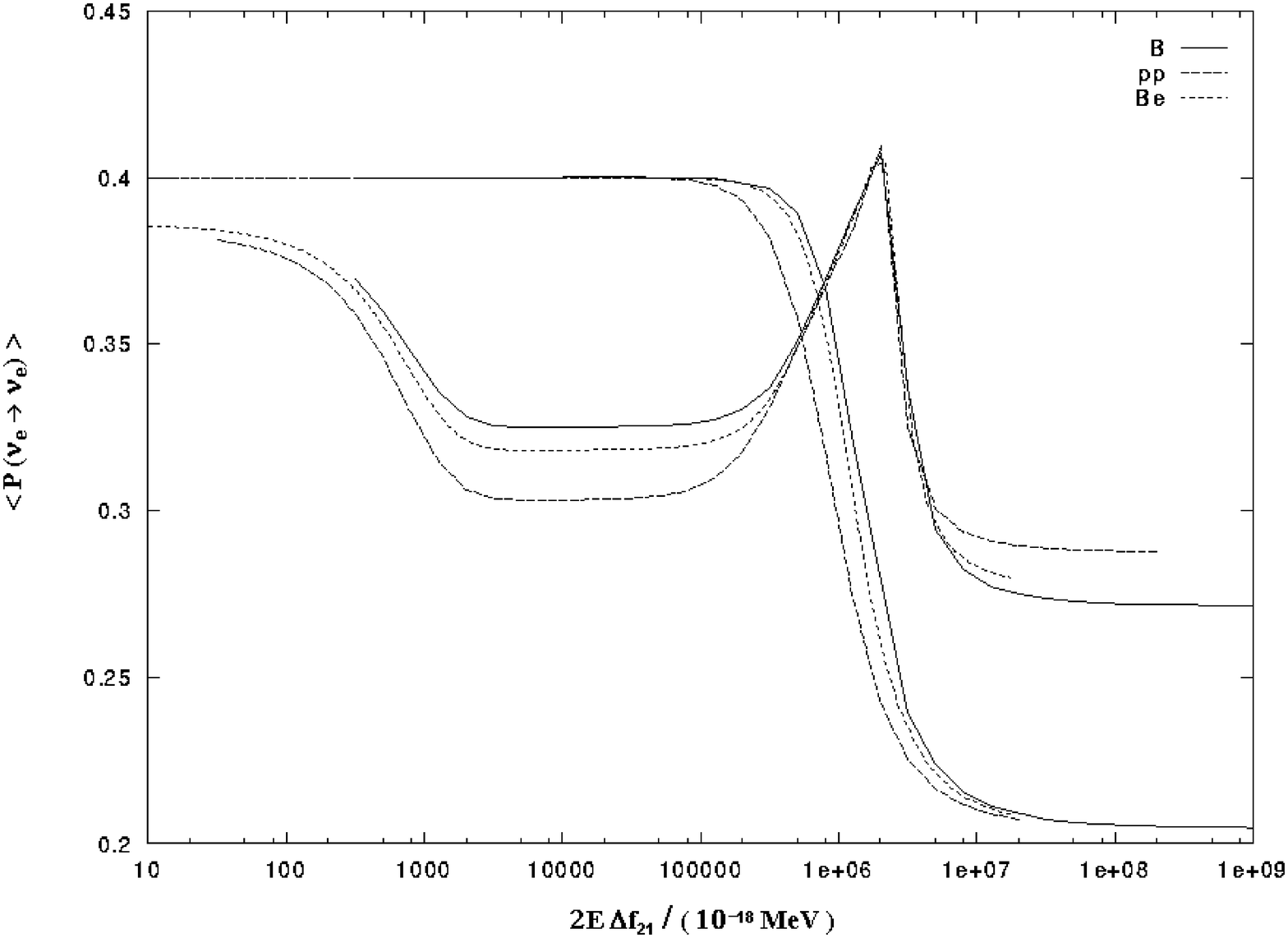}
\end{center}
\caption{$\Psurv$ for $\B$ and pp neutrinos, same values of $2E\df_{31}$ 
(``tub''--shaped curve is $2\times 10^6$).  $s_{2\th_{12}}=0.8~;~s_{13}^2 = 0.4$}
\label{lrglrg}
\end{figure}

\begin{figure}[h]
\begin{center}
\leavevmode
\epsfysize=290pt
\epsfbox[-145 73 753 725] {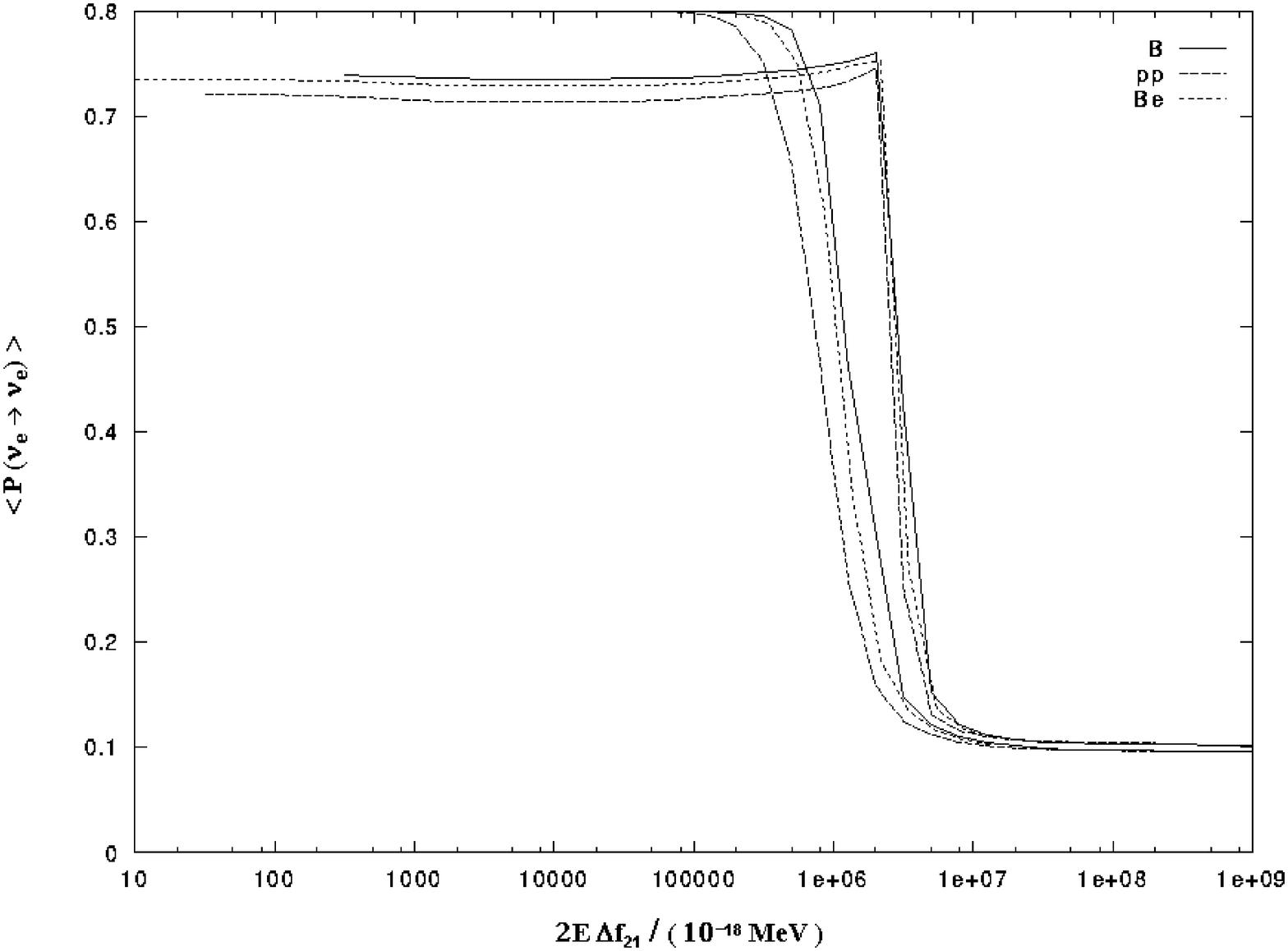}
\end{center}
\caption{$\Psurv$ for $\B$, $\Be$ and pp neutrinos, $2E\df_{31} = 2\times 10^4$ 
(left) and  $2\times 10^6$ (right).  $s_{2\th12}^2 = 0.8~;~s_{13}^2 = 0.8$.}
\label{big13}
\end{figure}

\begin{figure}[h]
\leavevmode
\epsfysize=300pt
\epsfbox[-80 80 740 710] {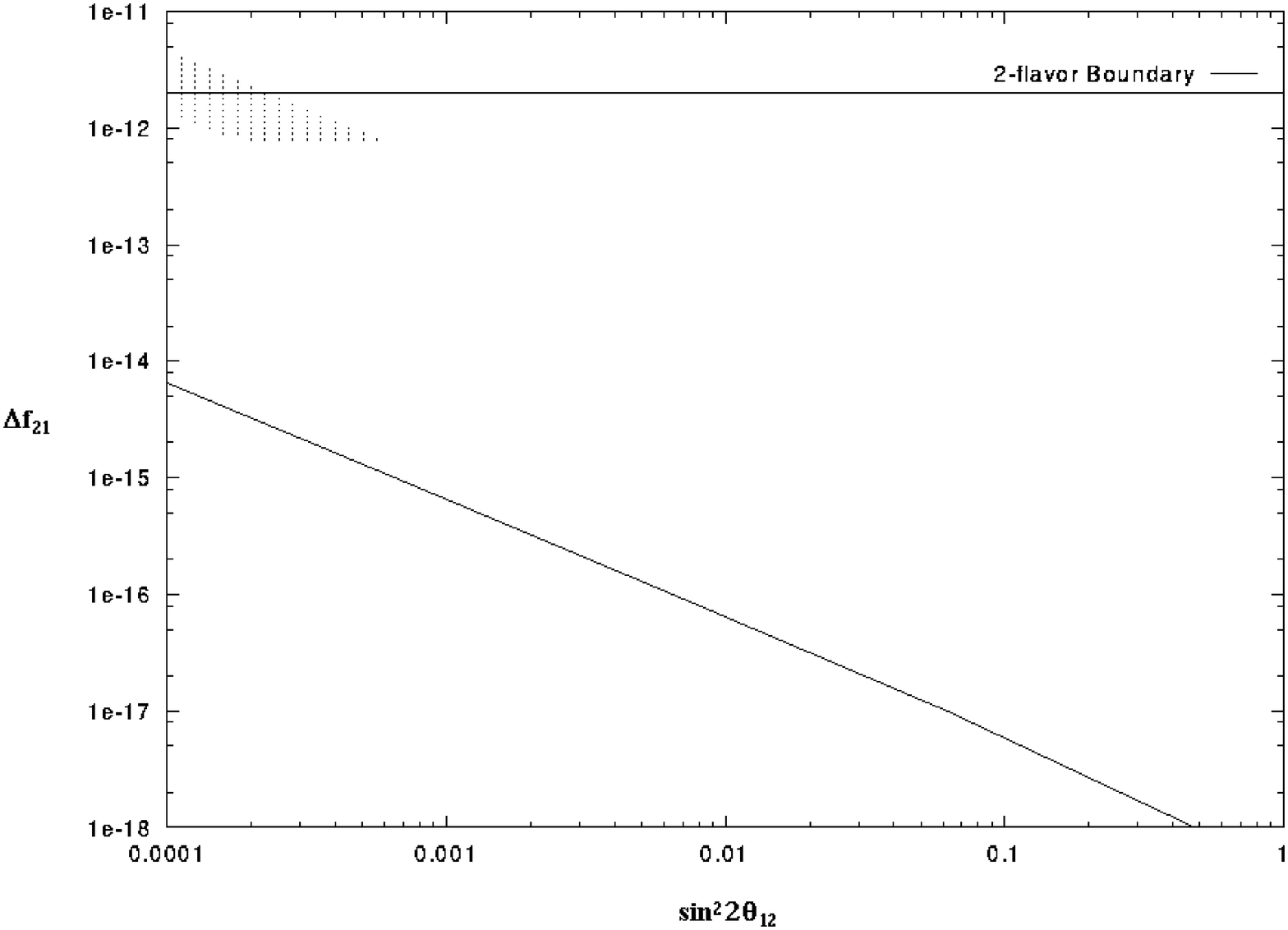}
\caption{$\df_{31} = 10^{-13}~;~s_{13}^2 = 10^{-3}~,~3\sigma$ C.L.}
\label{3f13s001}
\end{figure}

\begin{figure}[h]
\leavevmode
\epsfysize=300pt
\epsfbox[-80 80 740 710] {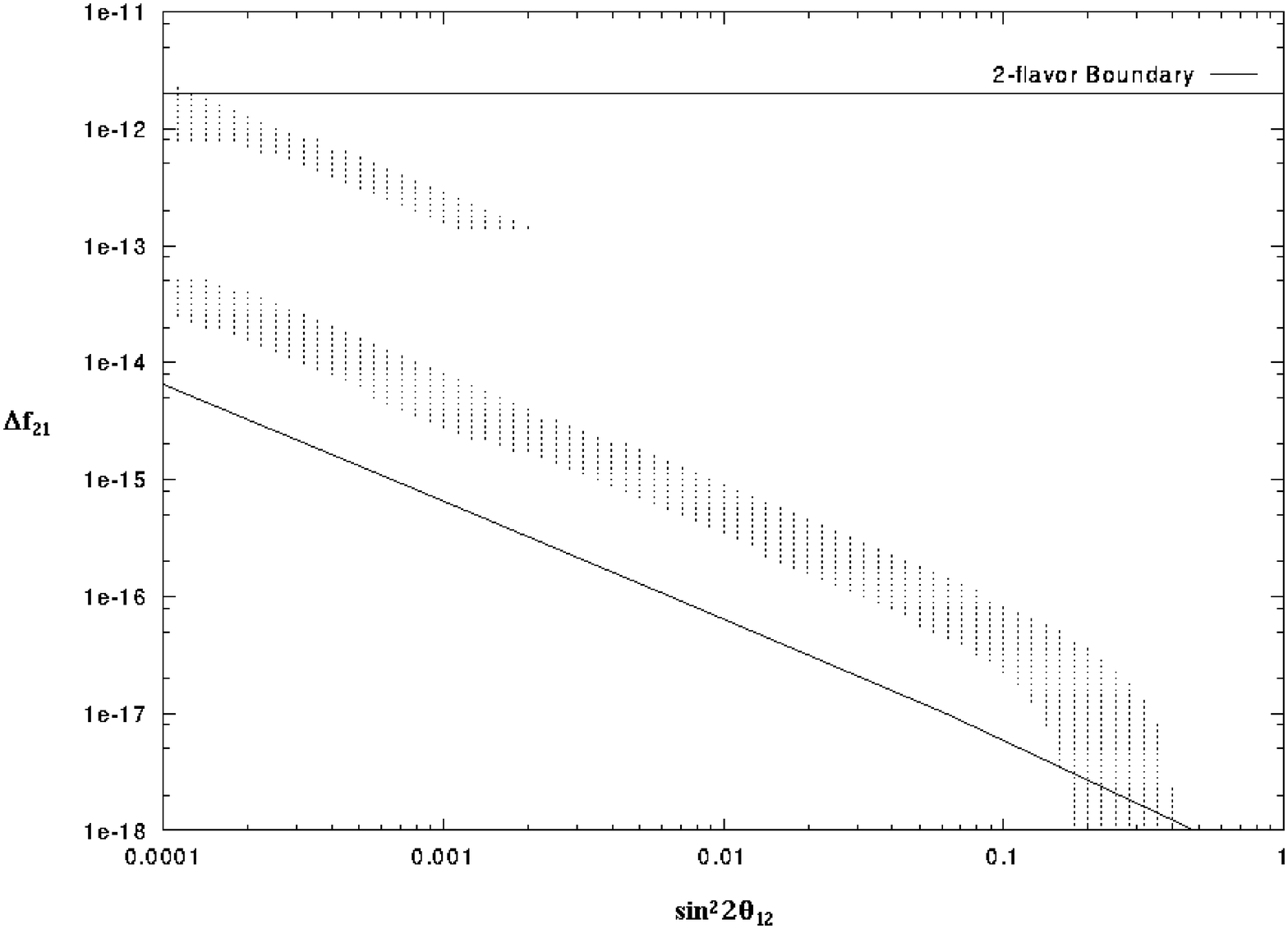}
\caption{Upper flux limit, $\df_{31} = 10^{-13}~;~s_{13}^2=10^{-3}~,
~3\sigma$ C.L.}
\label{u3f13s001}
\end{figure}

\begin{figure}[h]
\leavevmode
\epsfysize=300pt
\epsfbox[-80 80 740 710] {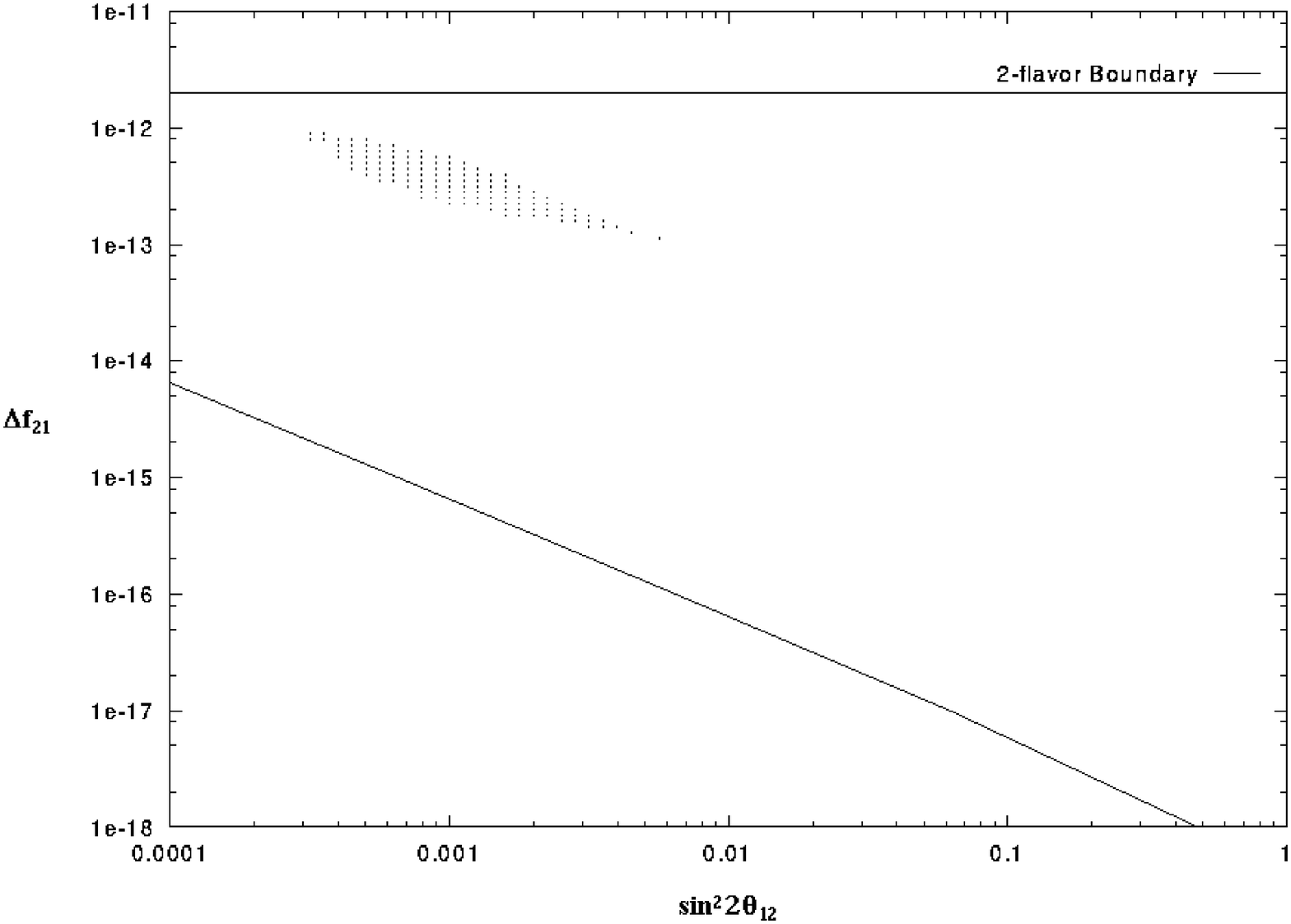}
\caption{$\df_{31} = 10^{-13}~;~s_{13}^2 = 10^{-4}~,~3\sigma$ C.L.}
\label{3f13s0001}
\end{figure}

\begin{figure}[h]
\leavevmode
\epsfysize=300pt
\epsfbox[-80 80 740 710] {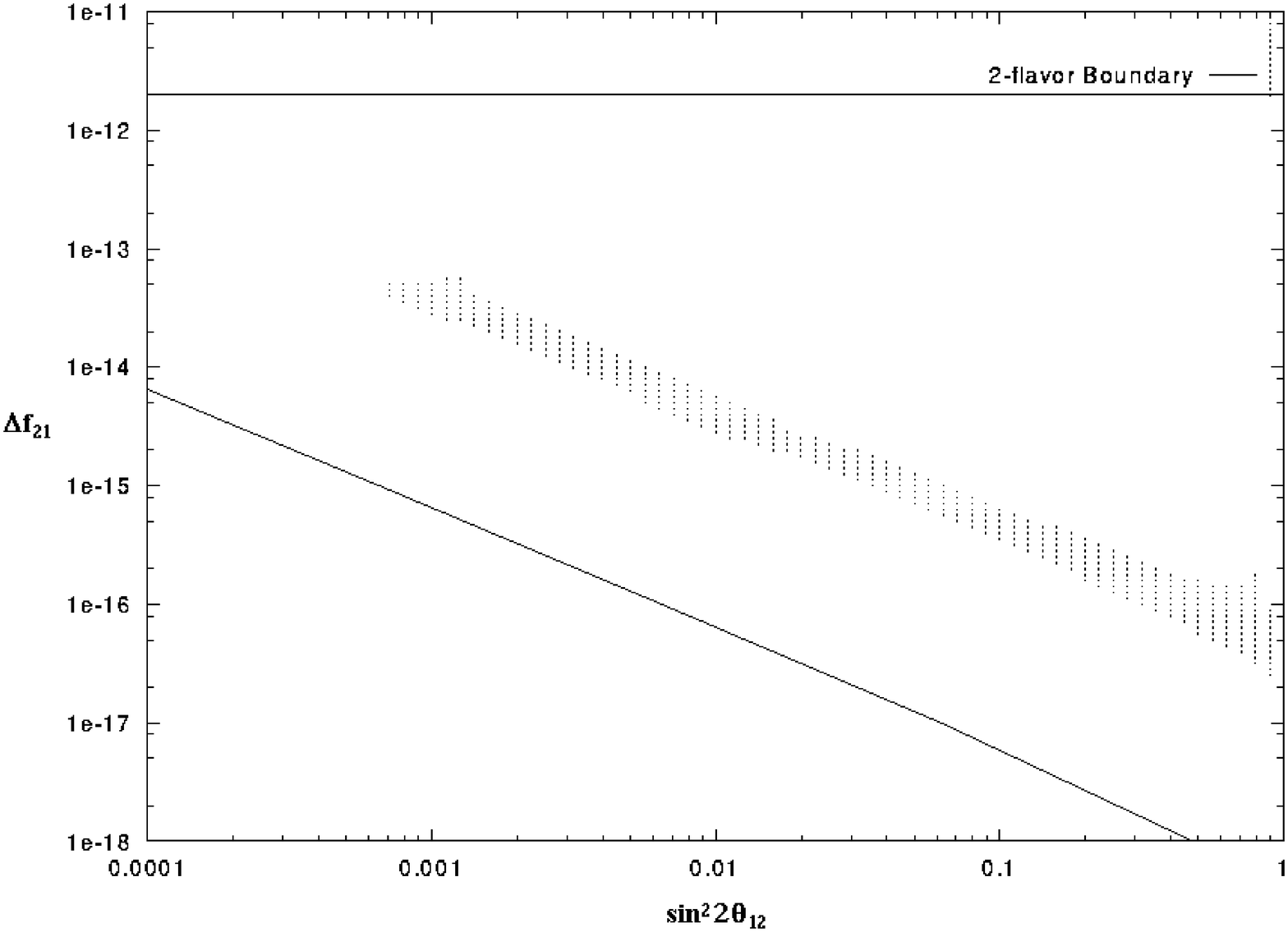}
\caption{$\df_{31} = 10^{-12}~;~s_{13}^2 = 10^{-4}~,~3\sigma$ C.L.}
\label{3f12s0001}
\end{figure}


\begin{figure}[h]
\leavevmode
\epsfysize=300pt
\epsfbox[-80 80 740 710] {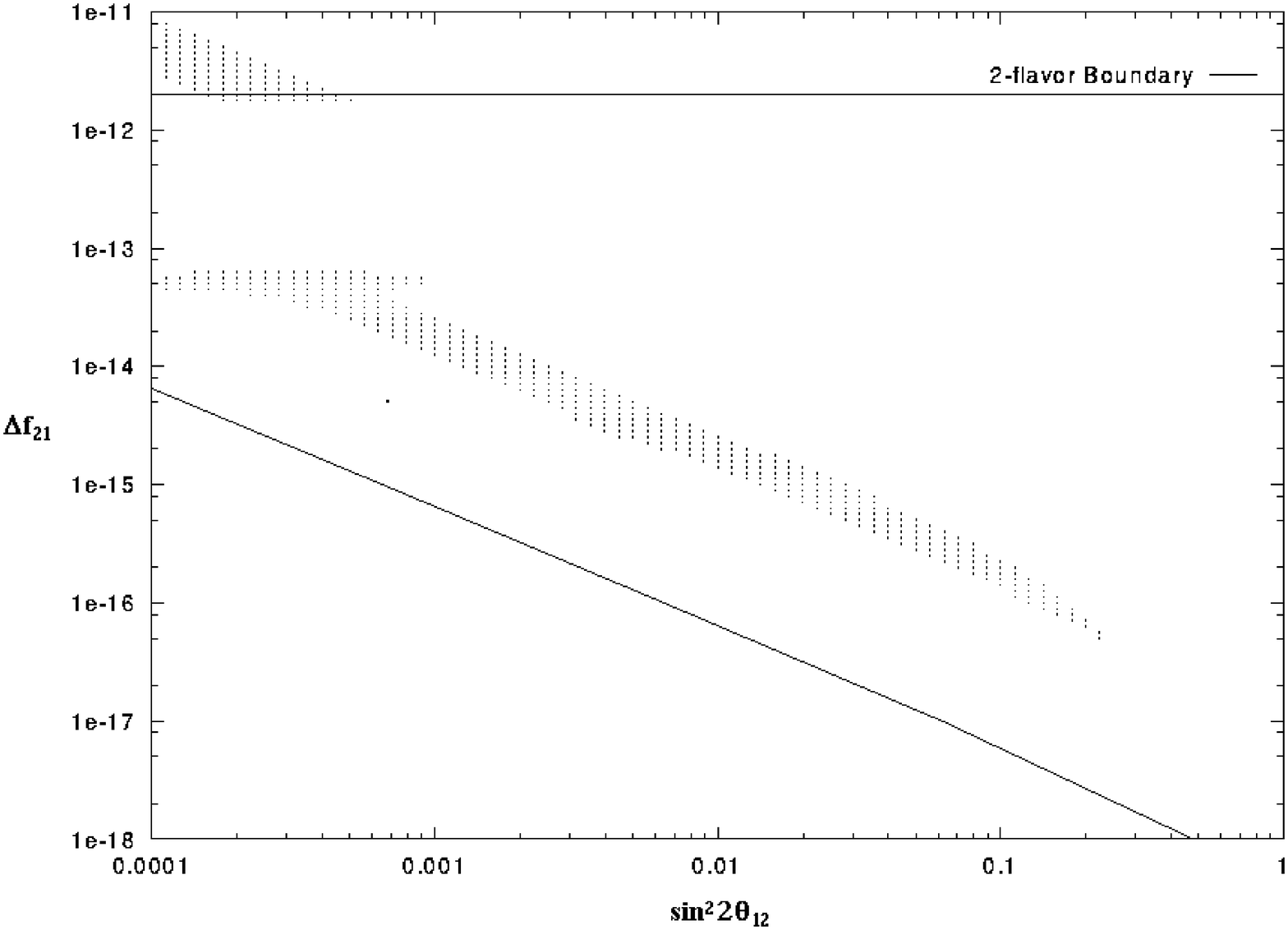}
\caption{$\df_{31} = 10^{-12}~;~s_{13}^2 = 0.2~,~3\sigma$ C.L.}
\label{3f12s2}
\end{figure}

\begin{figure}[h]
\leavevmode
\epsfysize=300pt
\epsfbox[-80 80 740 710] {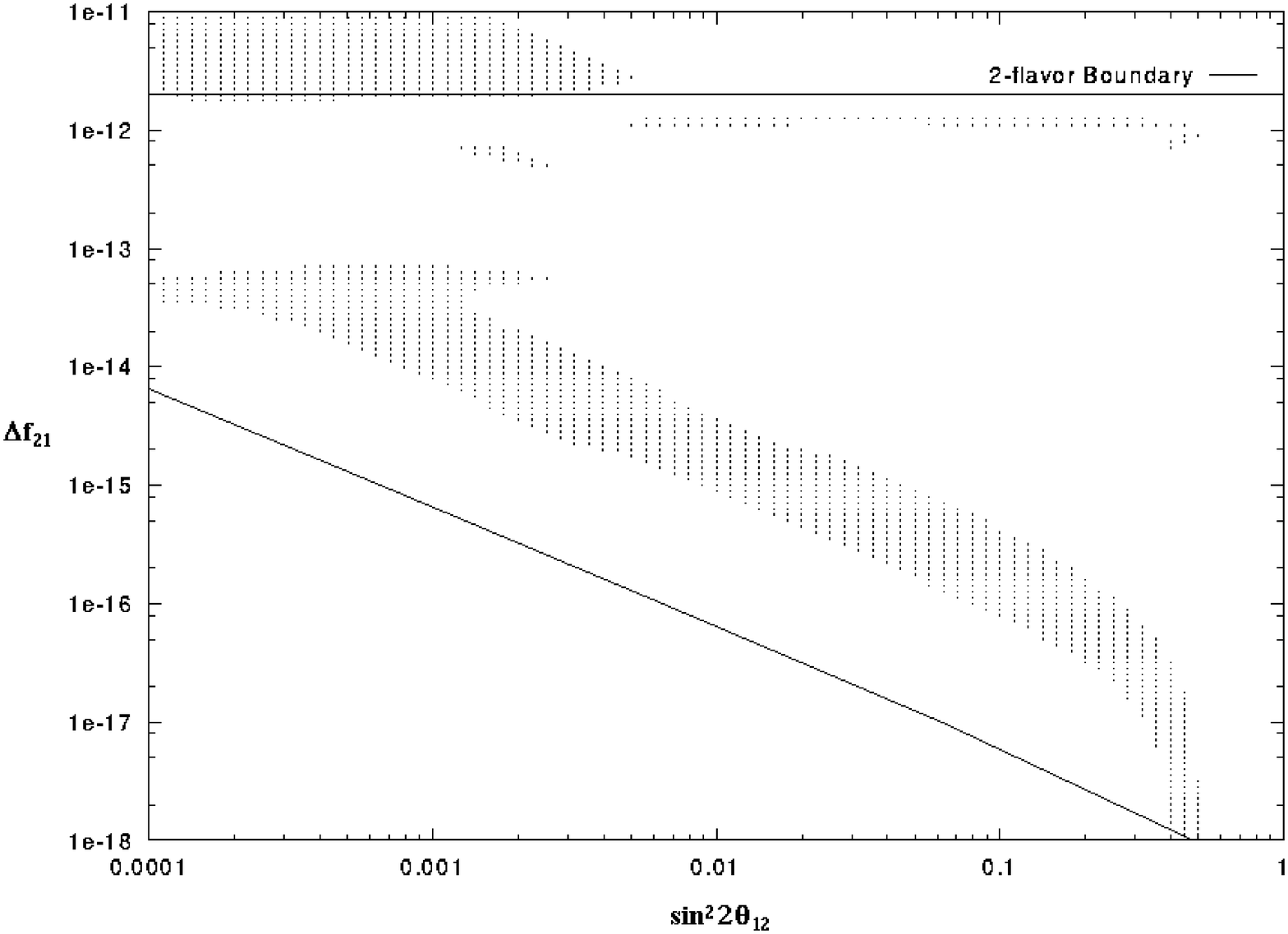}
\caption{$\df_{31} = 10^{-12}~;~s_{13}^2 = 0.3~,~3\sigma$ C.L.}
\label{3f12s3}
\end{figure}

\begin{figure}[h]
\leavevmode
\epsfysize=300pt
\epsfbox[-80 80 740 710] {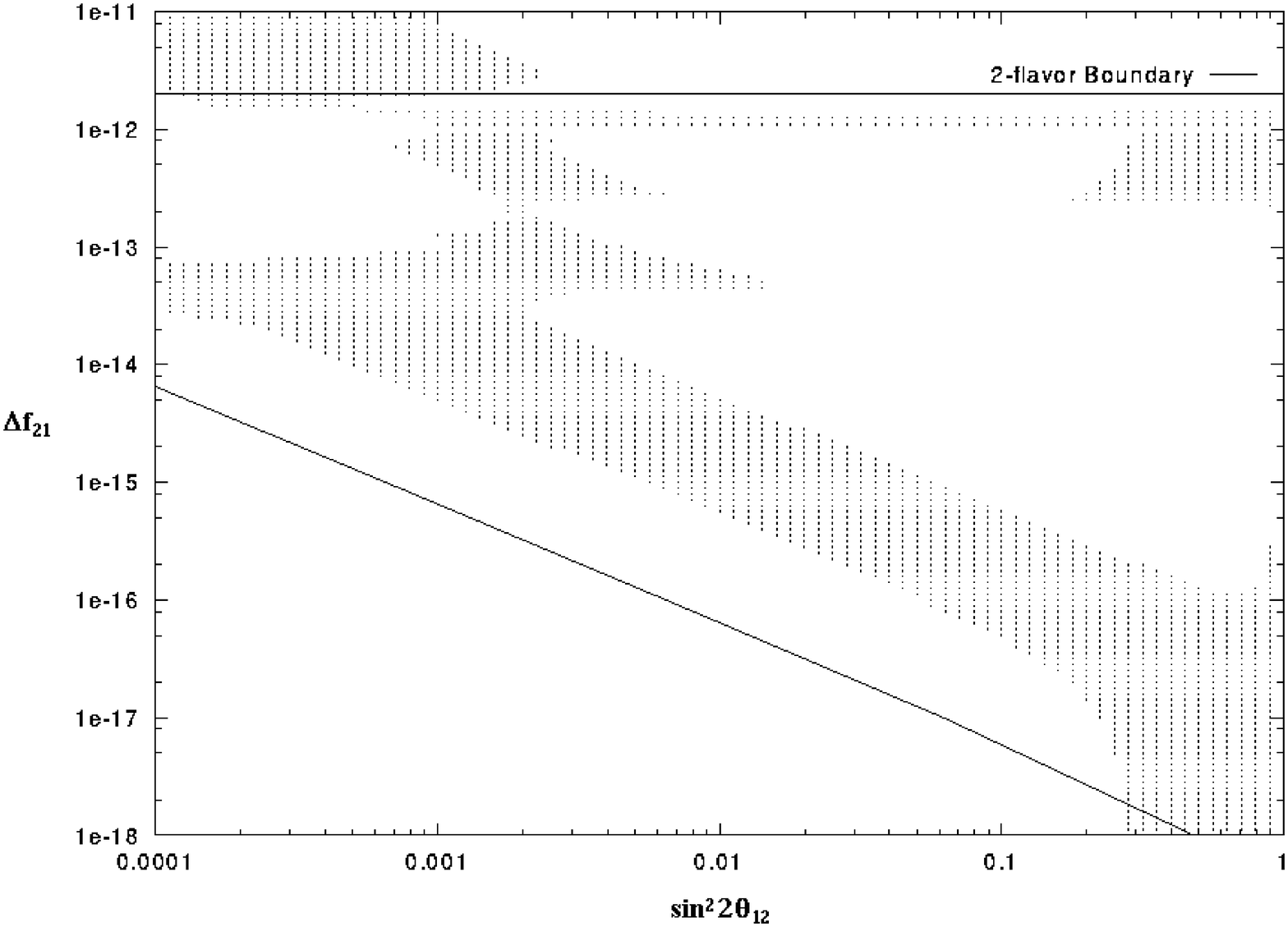}
\caption{$\df_{31} = 10^{-12}~;~s_{13}^2 = 0.4~,~3\sigma$ C.L.}
\label{3f12s4}
\end{figure}

\begin{figure}[h]
\leavevmode
\epsfysize=300pt
\epsfbox[-80 80 740 710] {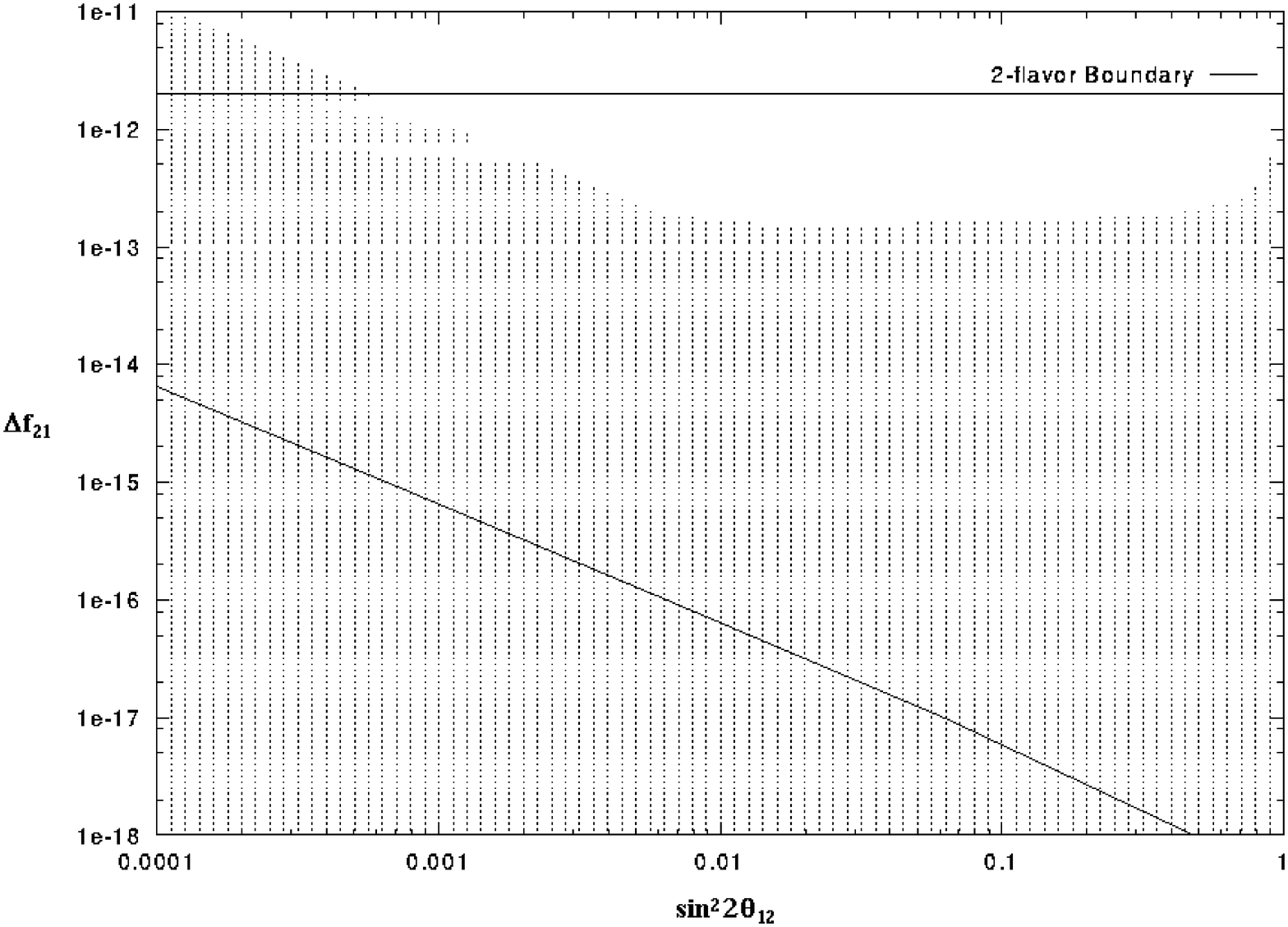}
\caption{$\df_{31} = 10^{-13}~;~s_{13}^2 = 0.4~,~3\sigma$ C.L.}
\label{3f13s4}
\end{figure}

\begin{figure}[h]
\leavevmode
\epsfysize=300pt
\epsfbox[-80 80 740 710] {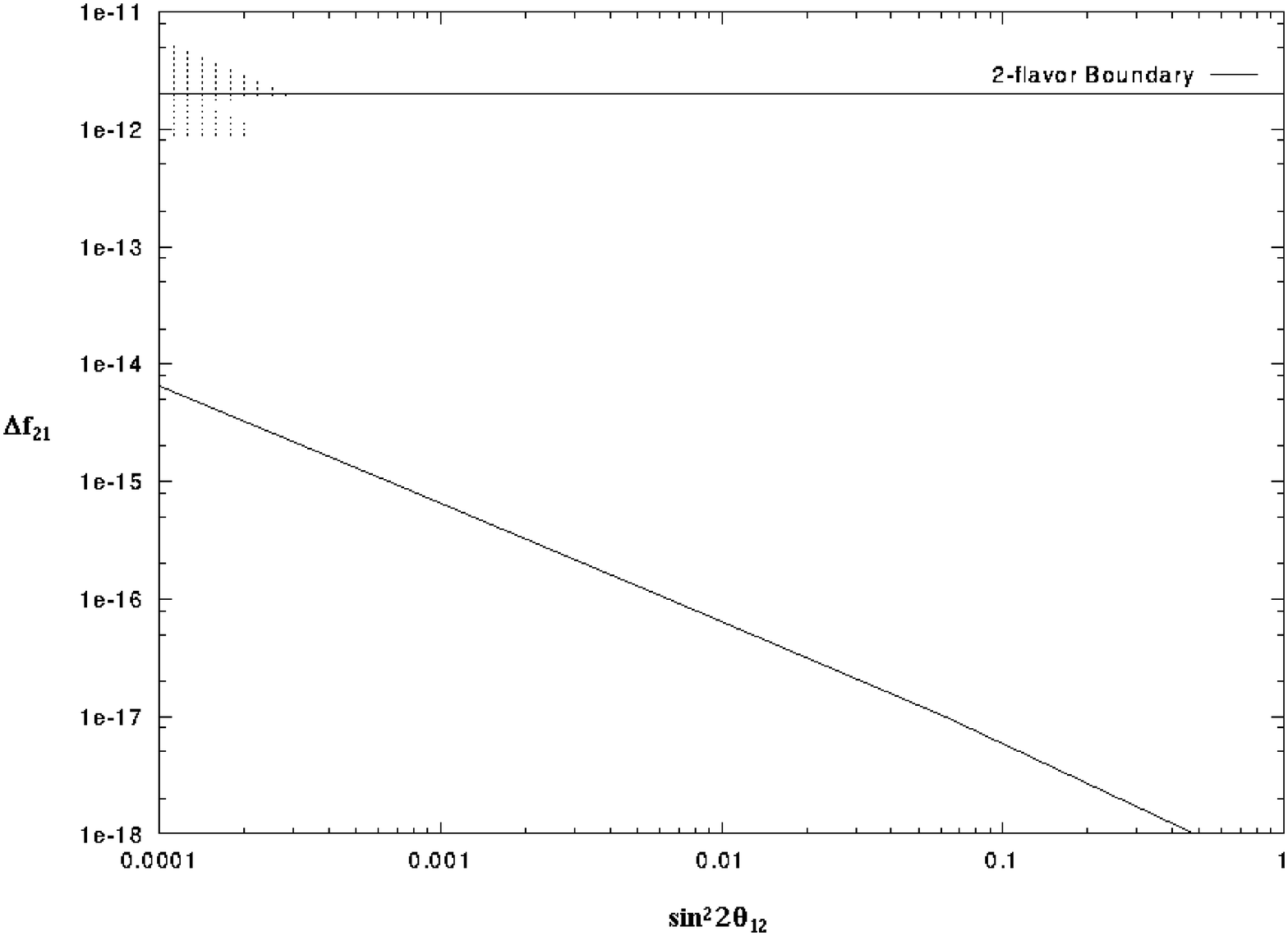}
\caption{$\df_{31} = 10^{-13}~;~s_{13}^2 = 0.4~,~2\sigma$ C.L.}
\label{2f13s4}
\end{figure}

\begin{figure}[h]
\leavevmode
\epsfysize=300pt
\epsfbox[-80 80 740 710] {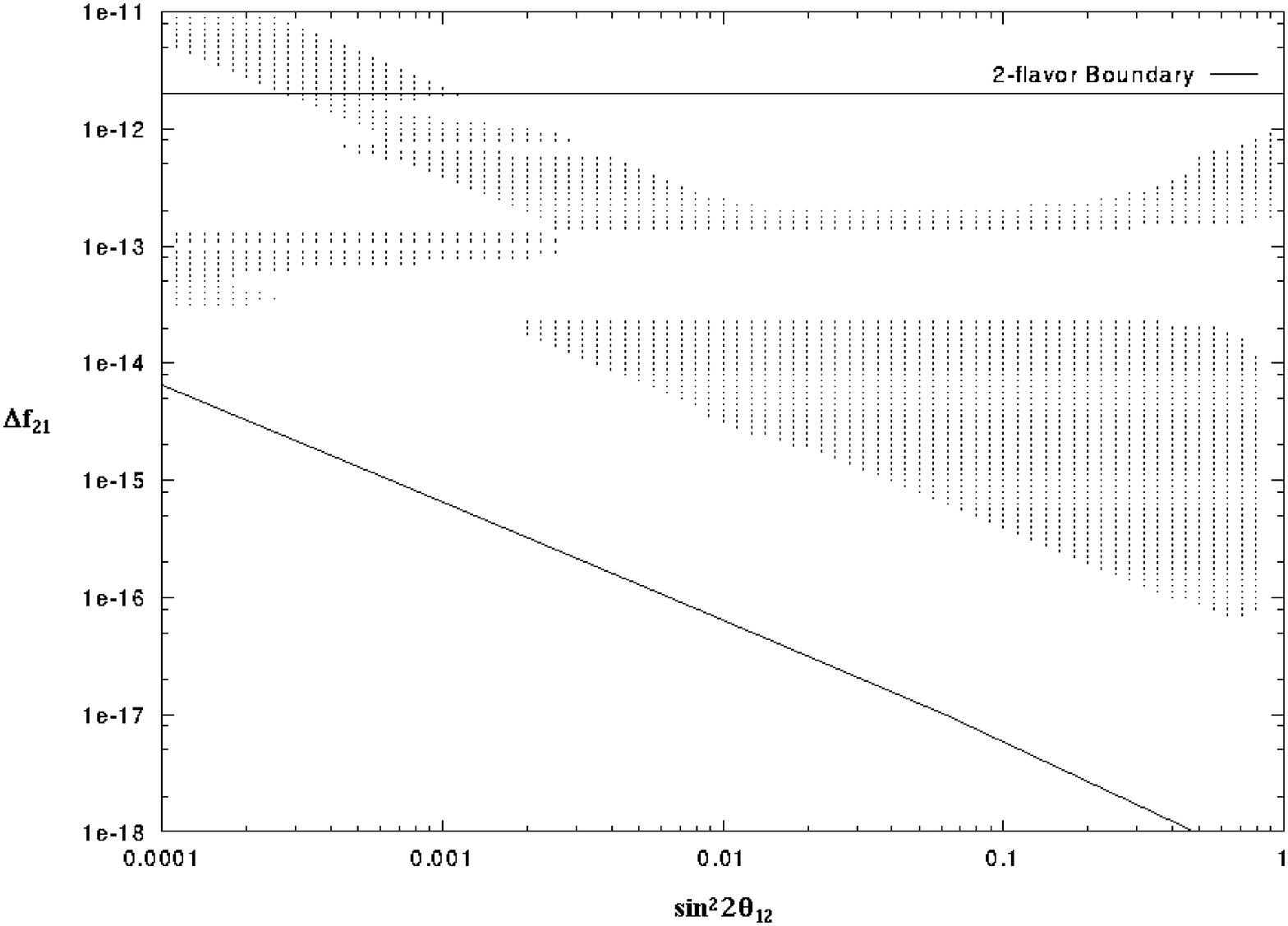}
\caption{Upper flux limit, $\df_{31} = 10^{-13}~;~s_{13}^2 = 
0.4~,~3\sigma$ C.L.}
\label{u3f13s4}
\end{figure}

\end{document}